\documentclass{amsart}
\usepackage{amssymb}
\usepackage{amsmath}
\usepackage{graphicx}
\usepackage[english]{babel}
\usepackage[pdftitle={HMM Spin},
  pdfauthor={Doghonay, Runborg},
  pdffitwindow=true,
  breaklinks=true,
  colorlinks=true,
  urlcolor=blue,
  linkcolor=red,
  citecolor=red,
  anchorcolor=red]{hyperref}
\usepackage[numbers]{natbib}

\numberwithin{equation}{section}
\numberwithin{table}{section}
\numberwithin{figure}{section}

\usepackage{graphicx}
\usepackage{amsfonts}
\usepackage{amsmath}

\newcommand{\vectorn}[1]{{ \bf{#1} }}

\newcommand{\unit}[1]{\ensuremath{ \,\mathrm{#1} }}
\newcommand{\dg}{\ensuremath{ {^\circ} }}
\newcommand{\dif}{\ensuremath{ \mathrm{d} }}
\newcommand{\dlt}{\ensuremath{ \mathrm{\Delta} }}
\newcommand{\e}{\varepsilon}

\newsavebox{\astrutbox}
\sbox{\astrutbox}{\rule[-5pt]{0pt}{20pt}}

\newtheorem{Remark}{Remark}

\begin{document}

\title[Atomistic-continuum multiscale modelling of magnetisation dynamics]{Atomistic-continuum multiscale modelling of magnetisation dynamics at non-zero temperature}

\author[D.~Arjmand]{Doghonay Arjmand}
\author[M.~Poluektov]{Mikhail Poluektov}
\author[G.~Kreiss]{Gunilla Kreiss} 
\email{doghonay.arjmand@it.uu.se,m.poluektov@outlook.com,gunilla.kreiss@it.uu.se}

\address{Division of Scientific Computing \\
  Department of Information Technology \\
  Uppsala University \\
  SE-751 05 Uppsala, Sweden.}

 \address{International Institute for Nanocomposites Manufacturing \\ 
 WMG, University of Warwick \\ 
 Coventry CV4 7AL, UK}

\address{Division of Scientific Computing \\
  Department of Information Technology \\
  Uppsala University \\
  SE-751 05 Uppsala, Sweden.}

%\subjclass[2010]{35B27, 65L12, 74Q10}

% *** not so easy!
% 65-XX Numerical Analysis
% 65Lxx	Ordinary differential equations
% 65L04  	Stiff equations
% 65L05  	Initial value problems
% 65L70  	Error bounds

\keywords{multiscale modelling, micromagnetism, atomistic-continuum coupling, Landau-Liftshitz-Gilbert equation}

%\date{\today}

\begin{abstract}
  In this article, a few problems related to multiscale modelling of magnetic materials at finite temperatures and possible ways of solving these problems are discussed. The discussion is mainly centred around two established multiscale concepts: the partitioned domain and the upscaling-based methodologies. The major challenge for both multiscale methods is to capture the correct value of magnetisation length accurately, which is affected by a random temperature-dependent force. Moreover, general limitations of these multiscale techniques in application to spin systems are discussed. 
\end{abstract}

\maketitle

\section{Introduction}
\label{sec:intro}

In recent years, atomistic spin dynamics (ASD) became an essential tool for understanding experimentally observed behaviour of magnetic materials \cite{Eriksson2016}. In this approach, the dynamics of spin magnetic moments of individual atoms is modelled. In contrast to ASD, the classical approach of understanding ferromagnets, which originates from the first half of twentieth century, is micromagnetics  \cite{Brown1963}. It operates with the volume-averaged quantities, such as magnetisation. The area of application of these two descriptions is somewhat different: ASD aims at describing the phenomena at the nanometre scale, while the micromagnetic theory is usually applicable at scales larger than the micrometre. Due to relatively high computational cost of ASD, it is useful to construct multiscale methods that combine ASD and micromagnetics in a single unified framework. An extended background and discussion of the purposes of such multiscale models can be found in \cite{Arjmand2016,Poluektov2016b}.

At finite temperatures, the rotational motion of the spin magnetic moments, which have the constant length, is described by a system of Langevin equations. The presence of temperature is modelled by a Gaussian noise term. In this case, each individual computational realisation is affected by a random force. However, the physically meaningful quantities of interest are the statistical averages of the magnetic moments, which have reduced lengths due to the cancellations in the averaging. For an accurate description of the dynamics, the reduction in the magnitude of the magnetisation or spin magnetic moments must be carefully accounted for. A traditional way of computing these statistical averages is the Monte Carlo method, see e.g. \cite{Landau2005}, which is computationally expensive, as large number of replicas of the solution is required to compute statistically meaningful quantities. An alternative way is to derive explicit equations for the evolution of statistical averages of the spin magnetic moments. One such approach is to use the Landau-Lifshitz-Bloch (LLB) equation, which describes the dynamics of the statistical averages, see e.g. \cite{Garanin1997}. A similar technique (a variant of the LLB) is proposed in \cite{Banas_Prohl2014} where the length (given by the so-called power law) is embedded into a macroscopic equation similar to LLB. These approaches typically contain some restrictive closure arguments, such as $\mathbb{E}[\vectorn{m} \times \vectorn{m}] = \mathbb{E}[\vectorn{m}] \times \mathbb{E}[\vectorn{m}]$, where $\mathbb{E}$ denotes an ensemble average and $\vectorn{m}$ is the spin magnetic moment, which are not necessarily true in general \cite{Tranchida2016}. To avoid such restrictive arguments, one may either resort to fully atomistic simulations, or design multiscale numerical algorithms which do not use these assumptions.

One of the most common types of concurrent multiscale methods is the \emph{partitioned domain} approach \cite{Tadmor2011}. Within this method, the computational domain is split into several regions with  different material descriptions, e.g. atomistic and continuum descriptions \cite{Miller2009}. The biggest challenge for these approaches is the construction of an error-free coupling at the interface separating the domains. There are several initial attempts of using partitioned domain atomistic-continuum multiscale models in application to micromagnetics \cite{GarciaSanchez2005,Jourdan2008,Andreas2014,Poluektov2016b}. 

Another common type of multiscale modelling is based on \emph{upscaling}, where the microscopic scale effects are systematically upscaled to an initially incomplete macroscopic model describing the macroscopic scale behaviour. In this case, the microscopic description has to be consistent with the current macroscopic variables, which leads to a two-way coupling between the microscale and macroscale models. There are several multiscale frameworks, such as the heterogeneous multicale method (HMM) \cite{Abdulle_E_Engquist2012,E_Engquist2003} or the equation-free approaches \cite{Kevrikidis_2003}, which use this idea. The upscaling-based strategy\footnote{Here, the term ``upscaling-based strategies'' is used to refer to the class of methods such as HMM or the equation-free methods, where a two-way coupling between the micro and macro scales is necessary for an accurate description of the macroscopic phenomena.} that is used in this paper rests upon the HMM methodology, where the microscale model is typically known, but expensive to solve over the entire macroscopic domain of interest, while the macroscale model is incomplete, as it lacks certain macroscopic quantities, which might be time-dependent. These missing quantities are then computed by carrying out the microscopic simulations in small spatial and temporal domains (sometimes referred to as representative volume elements). The methodology is efficient in the case of problems, which possess an inherent scale separation. The efficiency of the method is due to the fact that the size of the microscopic domains may be chosen comparable to the smallest length scale present in the system.

The aim of the present study is to describe the magnetisation dynamics using efficient atomistic-continuum multiscale formalisms, in which number of degrees of freedom is significantly reduced in comparison to a full atomistic simulation. Moreover, the overall ambition is to design algorithms, which do not suffer from restrictive closure arguments, and, at the same time, capture the correct dynamic behaviour. The microscale model, considered in the present work, is a discrete system of atomistic particles interacting at a finite temperature. For the purpose of this study, it is assumed that an atomistic model provides the \emph{exact} material behaviour, although this may not necessarily be true in reality. In a multiscale method,  a microscale model has to be consistently coupled to a macroscale model, which is addressed in this work under two different multiscale frameworks; the domain-partitioning and the HMM. Another aim of this work is to demonstrate the shortcomings of the standard continuum approach for a number of specific cases, which motivates the necessity of the multiscale modelling.

% REMOVED SENTENCES: repetion
% To reflect the main ideas, the HMM framework is briefly explained below.
% Heterogeneous multiscale method (HMM) is a general strategy for treating multiscale and possibly multiphysics problems \cite{Abdulle_E_Engquist2012,E_Engquist2003}. The name heterogeneous stands for the fact that the mathematical descriptions at different scales may be of different nature. The models are consistently coupled to describe the macroscopic behaviour of a multiscale system.
% HMM has two main components: a micro- and a macroscale model. 

In Section \ref{sec:tms} of this paper, a number of issues in relation to atomistic-continuum transition at non-zero temperatures are discussed, and the domain partitioning approach is presented. In Section \ref{Upscaling_Subsection}, the detailed mathematical formulation of the HMM approach for the case of micromagnetics is given. Afterwards, in Section \ref{sec:methods}, various multiscale examples are considered.

\section{Towards multiscale modelling}
\label{sec:tms}

The consistency between atomistic and continuum descriptions is required in any type of multiscale coupling. However, the construction of a deterministic continuum model that corresponds to the stochastic atomistic description is a complicated problem. Several issues related to estimation of the macroscopic quantities of stochastic atomistic systems are highlighted in subsequent subsections. The problem of selecting an appropriate continuum model is further discussed in subsection \ref{sec:model_sel}.

\subsection{Modelling approaches}

In this subsection, atomistic and continuum modelling approaches are summarised. For simplicity, a 1D spatial arrangement is considered. Full descriptions can be found in \cite{Evans2014} and \cite{Aharoni1996}, respectively. In what follows, the vector quantities are always denoted by bold face letters, e.g. $\vectorn{m} \in \mathbb{R}^{3}$, and the Euclidian 2-norm is written as $|\vectorn{m}| = \sqrt{m_x^2 + m_y^2 + m_z^2}$.

\subsubsection{Atomistic spin dynamics}

In the atomistic approach, the dynamic behaviour of spin magnetic moments of individual atoms is described by the atomistic Landau-Lifshitz-Gilbert equation \cite{Bergqvist2013,Evans2014}:
\begin{equation}
  \frac{\dif}{\dif t} \vectorn{m}_i = -\beta_\mathrm{L} \vectorn{m}_i \times \vectorn{H}_i - \alpha_\mathrm{L} \vectorn{m}_i
  \times \left( \vectorn{m}_i \times \vectorn{H}_i \right) , \quad \left|\vectorn{m}_i\right| = 1,
  \label{eq:ASD_LLG}
\end{equation}
\begin{equation}
  \beta_\mathrm{L} = \frac{\gamma}{1+\lambda^2} , \quad \alpha_\mathrm{L} = \frac{\gamma\lambda}{1+\lambda^2} ,
  \label{eq:ASD_param}
\end{equation}
\begin{equation}
  \vectorn{H}_i = \frac{1}{\mu}\left( \sum_j J_{ij} \vectorn{m}_j \right) +
  \frac{1}{\mu} K_\mathrm{a} \vectorn{p}_\mathrm{a}\vectorn{p}_\mathrm{a} \cdot \vectorn{m}_i + \vectorn{H}_\mathrm{e} + \vectorn{h}_i ,
  \label{eq:ASD_H}
\end{equation}
where $\gamma$ is the gyromagnetic ratio, $\lambda$ is the phenomenological (Gilbert) damping constant, $\vectorn{m}_i$ is the direction of spin magnetic moment, $\mu$ is the length of spin magnetic moment, $J_{ij}$ are constants of Heisenberg exchange interaction between atoms $i$ and $j$, $K_\mathrm{a}$ is the anisotropy constant, $\vectorn{p}_\mathrm{a}$ is the anisotropy axis (uniaxial anisotropy case is considered) and $\vectorn{H}_\mathrm{e}$ is the external field. Thermal excitations are taken into account by adding a stationary stochastic field with the following statistical properties: 
\begin{equation}
  \left\langle h_{i\rho}\left(t\right) \right\rangle = 0 ,\quad
  \left\langle h_{i\rho}\left(t\right) h_{j\nu}\left(s\right) \right\rangle = 2 D \delta_{ij} \delta_{\rho\nu} \delta\left(t-s\right) ,
  \label{eq:ASD_noise}
\end{equation}
\begin{equation}
  D = k_\mathrm{B}T \frac{\lambda}{\mu\gamma} ,
\end{equation}
where $\rho$ and $\nu$ are the Cartesian coordinates of $\vectorn{h}_i$, $k_\mathrm{B}$ is the Boltzmann constant and $T$ is temperature. More details regarding the stochastic term can be found in \cite{Brown1963b}. Parameters $\lambda$, $\mu$, $J_{ij}$, $K_\mathrm{a}$ and $\vectorn{p}_\mathrm{a}$ can be computed from electronic structure calculations \cite{Eriksson2016}, and are considered to be constant for a certain material.

\subsubsection{Continuum description of magnetisation dynamics at zero temperature}
\label{sec:cont}

In this subsection, the most simple (zero-temperature) continuum formulation is presented, which is used in partitioned domain examples in Section \ref{sec:methods}. The HMM framework uses different formulation of the continuum model, see Section \ref{Upscaling_Subsection}.

At the continuum scale, the dynamics of the magnetisation is modelled by the continuum version of the Landau-Lifshitz-Gilbert (LLG) equation \cite{Aharoni1996,Cimrak2008}:
\begin{equation}
  \frac{\partial}{\partial t} \vectorn{M} = -\beta_\mathrm{L} \vectorn{M} \times \vectorn{H} - \alpha_\mathrm{L} \vectorn{M}
  \times \left( \vectorn{M} \times \vectorn{H} \right) , \quad \left|\vectorn{M}\right| = 1,
  \label{eq:C_LLG}
\end{equation}
\begin{equation}
  \vectorn{H} = \frac{1}{\mu}A_\mathrm{e} \frac{\partial^2}{\partial x^2} \vectorn{M} +
  \frac{1}{\mu}K_\mathrm{a}\vectorn{p}_\mathrm{a}\vectorn{p}_\mathrm{a} \cdot \vectorn{M} + \vectorn{H}_\mathrm{e} ,
  \label{eq:C_H}
\end{equation}
where $\vectorn{M}$ is the normalised magnetisation field and $\beta_\mathrm{L}$ and $\alpha_\mathrm{L}$ are equal to atomistic parameters in  \eqref{eq:ASD_LLG}. At zero temperature, exchange parameter can be obtained directly from the atomistic parameters:
\begin{equation}
  A_\mathrm{e} = \frac{1}{2} \sum_j J_{ij} {r_{ij}}^2 ,
  \label{eq:C_exch_A}
\end{equation}
where $r_{ij}$ is the distance between atoms $i$ and $j$, and the sum is over all atoms with which atom $i$ interacts (also $A_\mathrm{e}$ is assumed to be spatially constant in this equation). Since the anisotropy term is local, the same anisotropy parameters $K_\mathrm{a}$ and $\vectorn{p}_\mathrm{a}$ are used in the continuum and the atomistic equations. The accuracy of these choices is discussed in the following subsections.

At finite temperatures, the continuum model has to be modified. These modifications differ depending on the approach and are further discussed in Section \ref{sec:model_sel}.

The continuum magnetisation field $\vectorn{M}$ is supposed to be equal to some spatial average of atomistic $\vectorn{m}_i$. However, due to the nature of the LLG equation, unit-length vectors are used in the formulations above. In general, atomistic-continuum transition introduces an error to the solution, which is dependent on magnetisation gradient, for details see \cite{Poluektov2016b}. 

\subsection{Direct numerical simulations of macroscopic quanteties for a 1D-system, using the atomistic model}

In this subsection, averaged quantities of 1D atomistic spin systems are investigated. The results highlight some problems with the standard continuum approach, which cannot capture the behaviour of some 1D atomistic systems.

The parameters of the dynamic deterministic continuum model, which is presented above, can be constructed out of the parameters of the dynamic deterministic atomistic model, which results in a certain solution error that depends on the magnetisation gradient \cite{Poluektov2016b}. In this case, the first step towards building a consistent continuum model that describes material behaviour at finite temperatures is the replacement of the stochastic atomistic description with a deterministic atomistic description with \emph{effective}\footnote{Effective parameters are such parameters that when substituted into an alternative model, result in an alternative solution, which is close to (or exactly the same as) the original solution provided by the original mode. In the present context, this means that the ensemble average of the solutions given by the stochastic atomistic model is equal to the solution of the deterministic atomistic model, which includes effective parameters.} parameters, from which the continuum parameters can be determined using known relations, e.g. \eqref{eq:C_exch_A}. In this subsection, it is shown that this approach results in an inaccurate model: such effective parameters should depend not only on temperature, but also on the state of magnetisation and other properties of the system, such as boundary conditions.

At finite temperatures, the mean-field approach \cite{Aharoni1996} is usually used, in which it is assumed that the statistical averages of atomistic spins interact with the same exchange constant ($J_{ij}$) as atomistic spins at $0\unit{K}$, while the length of spins is decreased due to averaging \cite{Aharoni1996}, i.e. $\left|\vectorn{m}_i\right|$ becomes less than $1$. From a numerical point of view, this can be implemented by replacing $\vectorn{m}_i$ with $\vectorn{m}_i^* s_i$, where $\left|\vectorn{m}_i^*\right| = 1$. If this assumption had been always valid, the only transition from the stochastic atomistic model at finite $T$ to the deterministic atomistic model would have been estimation of $s_i$. However, $s_i$ depends on the properties of the atomistic system, which is shown below. In subsection \ref{sec:temper}, it is demonstrated that the mean-field approach can give quantitatively different results in comparison to a direct numerical simulation.

It was already shown elsewhere that the effective exchange parameter depends on temperature \cite{Atxitia2010}. In subsection \ref{sec:nonlinear}, it is shown that it also depends on the state of magnetisation, therefore it is impossible to estimate an effective exchange parameter that is valid in a general case even for a particular temperature and a particular system.

\subsubsection{Dependence of macroscopic quantities on system size}

\begin{figure}
  \begin{center}
    \includegraphics[scale=0.8]{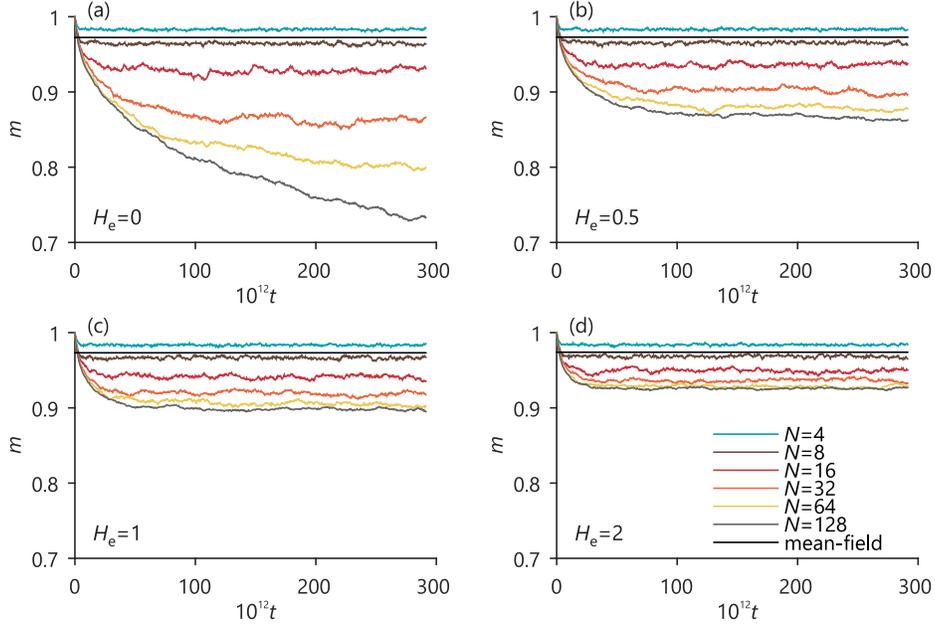}
  \end{center}
  \caption{Ensemble average of the particle-averaged spin magnetic moment of the system of $N$ atoms under external field depending on time. The magnitude of the external field, $H_\mathrm{e}$, is varied.}
  \label{fig:Asize}
\end{figure}

When material is modelled at an atomistic level, macroscopic quantities, which are extracted from the atomistic simulations, can depend on the size of the atomistic system. For example, in \cite{Kirschner2005}, the dependence of the magnetisation length, which was extracted from the atomistic simulations, on the size of the system was analysed. Such dependence exists for all types of boundary conditions, including periodic, but it is commonly accepted that there is a convergence with respect to the system size. In \cite{Kirschner2005}, in addition to material anisotropy, a weak external field was used to stabilise the atomistic system and align it in a certain direction. However, the presence of the external field has a far more profound effect on the macroscopic quantities, and there is a significant non-linear dependence of the magnetisation on the value of the external field. In Figure \ref{fig:Asize}, it can be seen that in the case of 1D arrangement of atoms (with periodic boundary conditions) not only the magnetisation length\footnote{Instead of a classically-defined magnetisation, which is a volume-average quantity, a particle-average quantity is used here, $m = \left| N^{-1} \mathbb{E} \left[ \sum \vectorn{m}_i \right] \right|$.} depends on the magnitude of the external field, but also the time required to reach the equilibrium. Moreover, when the external field is absent, convergence (within given time interval and for given range of system sizes) was not observed, although the material uniaxial anisotropy was taken into account. When an external field is present, convergence is observed.

\begin{table}
  \begin{center}
    \begin{tabular}{l*{5}{r}}
    \hline
    parameter         & $\mu$ & $J$ & $K_\mathrm{a}$ &
    $\vectorn{p}_\mathrm{a}$ & $\lambda$ \\
    \hline
    material 1        & $7.63\mu_\mathrm{B}$ & $1.28\cdot10^{-21}$ & $11.86\cdot10^{-24}$ &
    $\vectorn{e}_y$ & $0.1$ \\
    material 2        & $7.63\mu_\mathrm{B}$ & $1.28\cdot10^{-21}$ & $0$ &
    - & $0.1$ \\
    \hline
    \end{tabular}
  \end{center}
  \caption{Material parameters that were used in the simulations. Here, $\mu_\mathrm{B} = 9.274\cdot10^{-24}$ is the Bohr magneton.}
  \label{tab:sizeParam}
\end{table}

\begin{table}
  \begin{center}
    \begin{tabular}{l*{3}{r}}
    \hline
    parameter         & $\vectorn{H}_\mathrm{e}$ & $T$ & $N_\mathrm{s}$ \\
    \hline
    set 1        & $H_\mathrm{e}\vectorn{e}_y$ & $5$ & $100$ \\
    set 2        & \textbf{*} & $12$ & \textbf{**} \\
    \hline
    \end{tabular}\\
    {\textbf{*} - see Section \ref{sec:nonlinear}} \\
    {\textbf{**} - values are $3000$, $6000$ and $9000$ for $\phi = 135\dg$, $90\dg$ and $45\dg$ respectively} \\
  \end{center}
  \caption{Parameters that were used in the simulations. Here, $N_\mathrm{s}$ is the number of simulations used for ensemble averaging.}
  \label{tab:sizeParamSim}
\end{table}

Parameters that were used in the simulation are summarised in Table \ref{tab:sizeParam}, material 1, and Table \ref{tab:sizeParamSim}, set 1, and are taken from \cite{Evans2014}. Since this paper is methodology-oriented, units are ommitted for all quanitites, which is common in numerical analysis. Only nearest-neighbour interaction was taken into account. The implicit mid-point method \cite{dAquino2005} was used for time-stepping with a time step $\dlt t D = 0.005$. The initial directions of $\vectorn{m}_i$ were selected to be $\vectorn{e}_y$. The relatively low magnetisation value for this temperature can be explained by the 1D nature of the problem: each atom interacts only with two neighbours.

\subsubsection{Dependence of macroscopic quantities on temperature}
\label{sec:temper}

As discussed above, there is a strong dependence of the macroscopic quantities on temperature. There are several ways to quantify this dependence, e.g. the mean-field approach \cite{Aharoni1996} and the numerical modelling. As discussed in the introduction, within the analytical approach, a restrictive closure argument in the form of $\mathbb{E}[\vectorn{m} \times \vectorn{m}] = \mathbb{E}[\vectorn{m}] \times \mathbb{E}[\vectorn{m}]$ is made, which is not  valid in general \cite{Tranchida2016}. Moreover, due to its limitations, the analytical approach cannot provide either time required to reach the thermodynamic equilibrium or dependence of macroscopic quantities on the system size.

In Figure \ref{fig:Asize}, in addition to numerical results, analytical estimates are shown. Mean-field estimate of $m$ was obtained by solving the following equation \cite{Aharoni1996}:
\begin{equation}
  m = L \left( \frac{m}{k_\mathrm{B}T} \left( N_\mathrm{n} J + K_\mathrm{a} \right) + \frac{\mu H_\mathrm{e}}{k_\mathrm{B}T} \right) ,
  \label{eq:meanField}
\end{equation}
where $L$ is the Langevin function and $N_\mathrm{n}$ is the number of nearest neighbours, which in this case is $2$, see the appendix for a derivation of this formula in the presence of an external field and exchange interaction. The mean-field magnetisation length is significantly larger than the length of the magnetisation that is directly obtained from modelling of the dynamics of spin systems, as seen in Figure \ref{fig:Asize}. Moreover, the mean-field $m$ is less affected by the magnitude of the external field.

\subsubsection{Non-linear dependence of the effective exchange coefficient on the magnetisation gradient}
\label{sec:nonlinear}

In most cases, the macroscopic quantities of a stochastic spin system appear to be non-linearly dependent on external conditions. The non-linear dependence with respect to the external magnetic field was demonstrated in previous subsections. In this section, the importance of boundary conditions is highlighted. Specific boundary conditions are used here to create a magnetisation gradient, which influences effective exchange coefficient in a non-linear way. The approach is somewhat similar to the ``domain wall stiffness approach'' \cite{Atxitia2010,Hinzke2008}.

In the case of mechanical behaviour of 1D atomistic chain, a relevant macroscopic characteristic is the stiffness of the chain. It can be obtained numerically by applying forces to the boundary atoms and calculating the displacements of atoms. This method can be extrapolated to spin systems, where the interatomic interaction is Heisenberg exchange. An exchange stiffness of the entire 1D spin system is introduced. The spin system is twisted by the external magnetic field applied only to the boundary atoms.

\begin{figure}
  \begin{center}
    \includegraphics[scale=0.8]{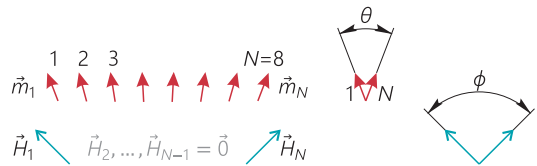}
  \end{center}
  \caption{Schematic illustration of the considered problem: atomistic chain of $8$ spins, with an external field applied only to the boundary atoms. The divergence angle of the external field vectors results in the constant gradient of spin directions created in the chain.}
  \label{fig:Afo_scheme}
\end{figure}

To illustrate the approach, the 1D chain of $N$ atoms interacting via Heisenberg exchange only with the nearest neighbours is considered. The system does not have magnetic anisotropy, $K_\mathrm{a}=0$; and open boundary conditions are used (atoms $1$ and $N$ interact only with one neighbour). External field $\vectorn{H}_\mathrm{e} = H \vectorn{e}_y$ is applied to atom $1$, external field $\vectorn{H}_\mathrm{e} = H \left( \vectorn{e}_x \sin\phi + \vectorn{e}_y \cos\phi \right)$ is applied to atom $N$, while external field $\vectorn{H}_\mathrm{e} = \vectorn{0}$ is applied to all other atoms. The problem is schematically illustrated in Figure \ref{fig:Afo_scheme}.

At $0\unit{K}$, the equilibrium configuration for this problem can easily be determined:
\begin{equation}
  \vectorn{m}_i = \vectorn{e}_x \sin\chi_i + \vectorn{e}_y \cos\chi_i , \quad
  \chi_i = \frac{\phi-\theta}{2} + \frac{i-1}{N-1}\theta,
  \label{eq:nlStiff_m}
\end{equation}
where $\theta$ is the solution of the following equation:
\begin{equation}   
  J \sin\frac{\theta}{N-1} = H \sin\frac{\phi-\theta}{2}.
  \label{eq:nlStiff}
\end{equation}
This means that there is a non-linear dependence between the magnitude of the applied field, $H$, the applied field divergence angle, $\phi$, and the spin moments divergence angle at the equilibrium, $\theta$.

At finite temperature, the equilibrium configuration can only be found by numerical simulation. After performing a number of simulations, it was observed that the orientation of the spin magnetic moments, $\chi_i$, of the ensemble average of the solution depends linearly on $i$, as well as in the deterministic case; however, the divergence angle, $\theta$ was different. Therefore, the behaviour of such material can be approximated by the deterministic model with a different (effective) parameter. When this deterministic model with an effective exchange constant $J^*$, is subjected to the same conditions as the stochastic model with the real exchange constant $J$, the resulting divergence angle of spin magnetic moments at the equilibrium, $\theta$, is the same. Hence the effective parameter $J^*$ can be determined by performing numerical simulations at finite temperature, extracting the angle $\theta$ from the ensemble average of the solution and using equation \eqref{eq:nlStiff}, in which $J$ is replaced by $J^*$.

\begin{figure}
  \begin{center}
    \includegraphics[scale=0.8]{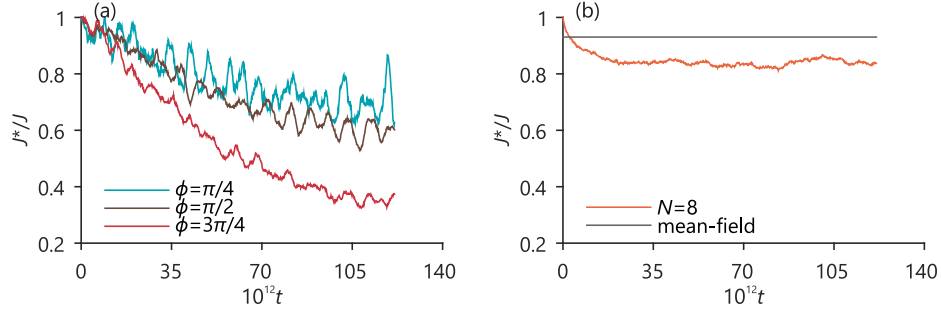}
  \end{center}
  \caption{The dependence of the normalised effective exchange coefficient of the constrained system on time (a) and the dependence of the particle-averaged spin magnetic moment of the unconstrained system on time (b).}
  \label{fig:Afo}
\end{figure}

In Figure \ref{fig:Afo}a, the effective exchange coefficient between two spins, $J^*$, normalised by $J$ is plotted. It is clearly seen that the effective exchange constant depends non-linearly on the angle between the magnetic fields applied to the boundary atoms. This implies a non-linear dependence of the continuum exchange constant on the local magnetisation gradient in the case of finite temperatures.

Problem parameters are summarised in Table \ref{tab:sizeParam}, material 2, and Table \ref{tab:sizeParamSim}, set 2. A time step $\dlt t D = 0.005$ was used. The initial directions of $\vectorn{m}_i$ were selected to be according to equation \eqref{eq:nlStiff_m} with real $J$, hence $J^* J^{-1} = 1$ at $t=0$. A chain of $8$ atoms was considered. Angle $\theta$ was determined by a linear fit of angles $\chi_i$, which are obtained by projecting the ensemble average of the solution on $xy$-plane. The following value of $H$ was selected:
\begin{equation}
  H = \frac{J \sin\frac{\phi}{2N-2}}{\sin\frac{\phi}{4}} ,
\end{equation}
which minimises the statistical error of calculation of $J^*$ for small divergence angles. The angle $\phi$ was varied: values of $135\dg$, $90\dg$ and $45\dg$ were used.

The coefficient $J^* J^{-1}$ shows the change of the exchange parameter when the constrained (i.e. with the external field applied to the boundary atoms) stochastic system is replaced by the constrained deterministic. In the case of unconstrained system (i.e. no external field), if the mean-field assumption is to be believed, the stochastic system can be replaced by the deterministic system with the same exchange parameter, but different magnetisation length. After re-normalisation $\vectorn{m}_i = \vectorn{m}_i^* s$, $\left|\vectorn{m}_i^*\right| = 1$, this replacement becomes mathematically equivalent to changing the exchange parameter, i.e. $s = J^* J^{-1}$. For an unconstrained system with the same number of atoms, see Figure \ref{fig:Afo}b, this coefficient is significantly larger than for the constrained system. This confirms that if greater accuracy is required, the effective parameters, which are used in the deterministic LLG equation, should not be pre-calculated constants (for a given temperature) but should be functions of the local magnetisation and the local magnetisation gradient. 

The simulation in Figure \ref{fig:Afo}b was performed with the same material parameters as used for the constrained system; however, without an external field; $N_\mathrm{s} = 100$ number of realisations  was used. The coefficient $s = J^* J^{-1}$ in this simulation was calculated as the length of particle-averaged spin magnetic moment. The mean-field value was obtained using equation \eqref{eq:meanField}.

\subsection{Partitioned domain approach}
\label{sec:part_dom}

The application of partitioned domain approach to the modelling of magnetisation dynamics is covered in \cite{Poluektov2016b,Poluektov2016}. The key idea behind the methodology is the separation of the computational domain into two regions: atomistic and continuum, which are connected at the interface. In the case of nearest-neighbour interatomic interaction, neither transition region, nor padding atoms (atoms behaviour of which are obtained by interpolation of the solution at neighbouring continuum nodes) are necessary. In the 1D case, the interface consists of a single point and the coupling is straightforward: the interface point interacts with the nearest continuum node as a continuum node and it interacts with the nearest atom as an atom (in the case of nearest-neighbour interatomic interaction). However, when the dynamic material behaviour is modelled, the introduction of an additional damping region at the interface is necessary \cite{ChubykaloFesenko2006}. This additional numerical damping is applied to the atoms in the proximity of the interface and acts as a low-pass filter that absorbs high-frequency waves that are otherwise reflected from the interface (due to the difference in discretisation parameters) and contribute to numerical error. The subsequent subsection follows the analogous section from \cite{Poluektov2016b}.  

\subsubsection{Damping band}

The equation describing the dynamics of numerically-damped atoms is the following \cite{Poluektov2016}:
\begin{equation}
  \frac{\dif}{\dif t} \vectorn{m}_i = -\beta_\mathrm{L} \vectorn{m}_i \times \vectorn{H}_i - \alpha_\mathrm{L} \vectorn{m}_i
  \times \left( \vectorn{m}_i \times \vectorn{H}_i \right) - \gamma_{\mathrm{D}i} \vectorn{m}_i \times \left( \vectorn{m}_i \times \vectorn{m}_{\mathrm{A}i} \right),
  \label{eq:dampBand_ASD_LLG}
\end{equation}
where $\gamma_{\mathrm{D}i}$ is the strength of the damping for atom $i$ and $\vectorn{m}_{\mathrm{A}i}$ is the normalised average of the solution over a certain region. The damping strength increases gradually with atomic positions approaching the interface and is proportional to the time derivative of $\vectorn{m}_i$:
\begin{equation}
  \gamma_{\mathrm{D}i} = g_\mathrm{D} \left( \frac{r_i^\mathrm{R}}{r_i^\mathrm{R}+r_i^\mathrm{P}-a} \right)^2 \sqrt{\left| \frac{\dif}{\dif t} \vectorn{m}_i \right|},
  \label{eq:dampBand_gamD}
\end{equation}
where $g_\mathrm{D}$ is a constant determining the damping strength, $r_i^\mathrm{R}$ is the distance to the closest atom outside of the damping band, $r_i^\mathrm{P}$ is the distance to the interface atom/node and $a$ is the interatomic spacing.

Within the damping band, the solution is attenuated to an average value, $\vectorn{m}_{\mathrm{A}i}$, which is calculated within a certain window surrounding atom $i$. The width of the window is denoted as $s_\mathrm{A}$. An additional average quantity $\vectorn{m}^*_{\mathrm{A}i}$ is introduced:
\begin{equation}
  \vectorn{m}^*_{\mathrm{A}i} = \vectorn{c}_2 x^2 +  \vectorn{c}_1 x + \vectorn{c}_0.
  \label{eq:dampBand_mAver}
\end{equation}
Coefficients $\vectorn{c}$ are fitted to all $\vectorn{m}_j$ within the averaging window using least squares. Finally, $\vectorn{m}_{\mathrm{A}i}$ is obtained by normalisation:
\begin{equation}
  \vectorn{m}_{\mathrm{A}i} = \frac{ \vectorn{m}^*_{\mathrm{A}i} }{ \left| \vectorn{m}^*_{\mathrm{A}i} \right| }.
  \label{eq:dampBand_mAver_norm}
\end{equation}

\subsection{Selection of the macroscopic scale model for the multiscale framework}
\label{sec:model_sel}

Due to the highlighted issues, the selection of an appropriate model for the macroscopic scale in the case of domain partitioning is ambiguous. First of all, based on the type of the problem, the nature of the macroscopic model has to be chosen, which can be quasistatic\footnote{Quasistatic models describe systems, evolution of which is slow enough for the system to remain in internal equilibrium. In the case of micromagnetism, quasistatic models are simply static equations (i.e. without time derivative of magnetisation), however with time-dependent boundary conditions.}, dynamic deterministic or dynamic stochastic.

Quasistatic models are suitable for the cases when an equilibrium magnetisation distribution around an isolated atomistic region, which can be fully dynamic, has to be determined. The example of such coupling for the case of mechanical behaviour (coupling molecular dynamics and elastic finite-element continuum descriptions) is the CADD method \cite{Qu2005}. Since it might be unphysical to model all the waves generated within the atomistic region, which eventually diffuse and are converted to heat over large distances, there is an additional advantage of using a quasistatic continuum model as it avoids such problems.

In cases when dynamic phenomena should be modelled within the continuum region, e.g. domain wall movement, one of the dynamic models has to be used. In this case, while waves are transmitted within the continuum, atomistic regions are idle and thereby reduce computational efficiency; however, this issue cannot be avoided for such type of problems. The most simple model, which can be used for the dynamic continuum, is the standard Landau-Lifshitz-Gilbert equation, as presented in section \ref{sec:cont}, with the length of magnetisation rescaled due to temperature dependency (note that $\left|\vectorn{M}\right|=1$, but parameters in front of $\vectorn{M}$ are rescaled). However, such an approach cannot describe a local change of the magnetisation length due to change of its orientation with respect to the external field or anisotropy direction, which might be important for some applications. This problem can be handled partly by assuming space- and time-dependence of the magnetisation length, and adding an additional PDE describing its evolution. In this case, such PDE should be either derived from the atomistic equation or compiled based on empirical observations.

The last option for the continuum region is the stochastic Landau-Lifshitz-Bloch equation \cite{Garanin1997,ChubykaloFesenko2006}. In this approach, due to the stochastic nature and the structure of the dynamic equation, local changes of the expected value of magnetisation length are permitted. The advantage of the stochastic LLB is that it can naturally describe domain walls with nonconstant magnetisation length \cite{ChubykaloFesenko2006}. However, in other areas of physics, such as modelling the mechanical behaviour of solids, stochastic descriptions of materials are not typically used at the macroscopic scale \cite{Tadmor2011}, since stochastic effects diminish with the system size. 

\section{An upscaling-based approach}
\label{Upscaling_Subsection}

In subsection \ref{sec:model_sel}, various ways of selecting a macroscopic model and their limitations were mentioned. In this section, an alternative strategy based on HMM is proposed to compute approximate \emph{upscaled} macroscopic quantities (statistical averages of the atomistic spins given by \eqref{eq:ASD_LLG}). The approach should be seen as complementary to the domain partitioning approach that is described in subsection \ref{sec:part_dom}, whereby the continuum equations (needed for the domain partitioning) are obtained computationally. Before proceeding with the description of the method, few mathematical tools and notations are presented.

\subsubsection{Averaging kernels} 

A key ingredient of the HMM is a compression step, which typically uses some general purpose averaging kernels. To illustrate the idea, $f^{\e}(t) = f(t/\e)$ is assumed, where $f$ is  a $1$-periodic integrable function. A local average of this function can be computed by selecting a small region $I_{\eta} = [-\eta/2,\eta/2]$ at first and then averaging $f^{\e}$ over $I_{\eta}$ afterwards. This operation, however, results in the following error:
\begin{equation}
  \left| \dfrac{1}{\eta} \int_{-\eta/2}^{\eta/2} f(t/\e) \; \mathrm{d}t  - \int_{0}^{1} f(z) \;  \mathrm{d}z \right| \leq C \dfrac{\e}{\eta},
  \quad \text{for some } \eta > \e. 
  \label{eq_NaiveAveraging_Estimate}
\end{equation}
This averaging can be done more accurately using the space of averaging kernels $\mathbb{K}^{p,q}$ which consists of symmetric functions $K$, such that $K(t) = K(-t)$, $K$ is compactly supported in $[-1/2,1/2]$, and $\frac{\mathrm{d}^{(q+1)} K }{\mathrm{d} t^{q+1}} \in BV(\mathbb{R})$, where $BV$ is the space of functions with bounded variations in $\mathbb{R}$, and the derivative is understood in the weak sense. Moreover, the parameter $p \in \mathbb{Z}^{+}$ represents the number of vanishing moments, i.e. $K$ has $p$ vanishing moments if
\begin{equation*}
  \int_{\mathbb{R}} K(t) t^r \;  \mathrm{d}t  = 
  \begin{cases} 
  1, & r=0, \\
  0, & 0<r \leq p.
  \end{cases}
\end{equation*}

\begin{figure}
  \begin{center}
    \includegraphics[width=0.47\textwidth]{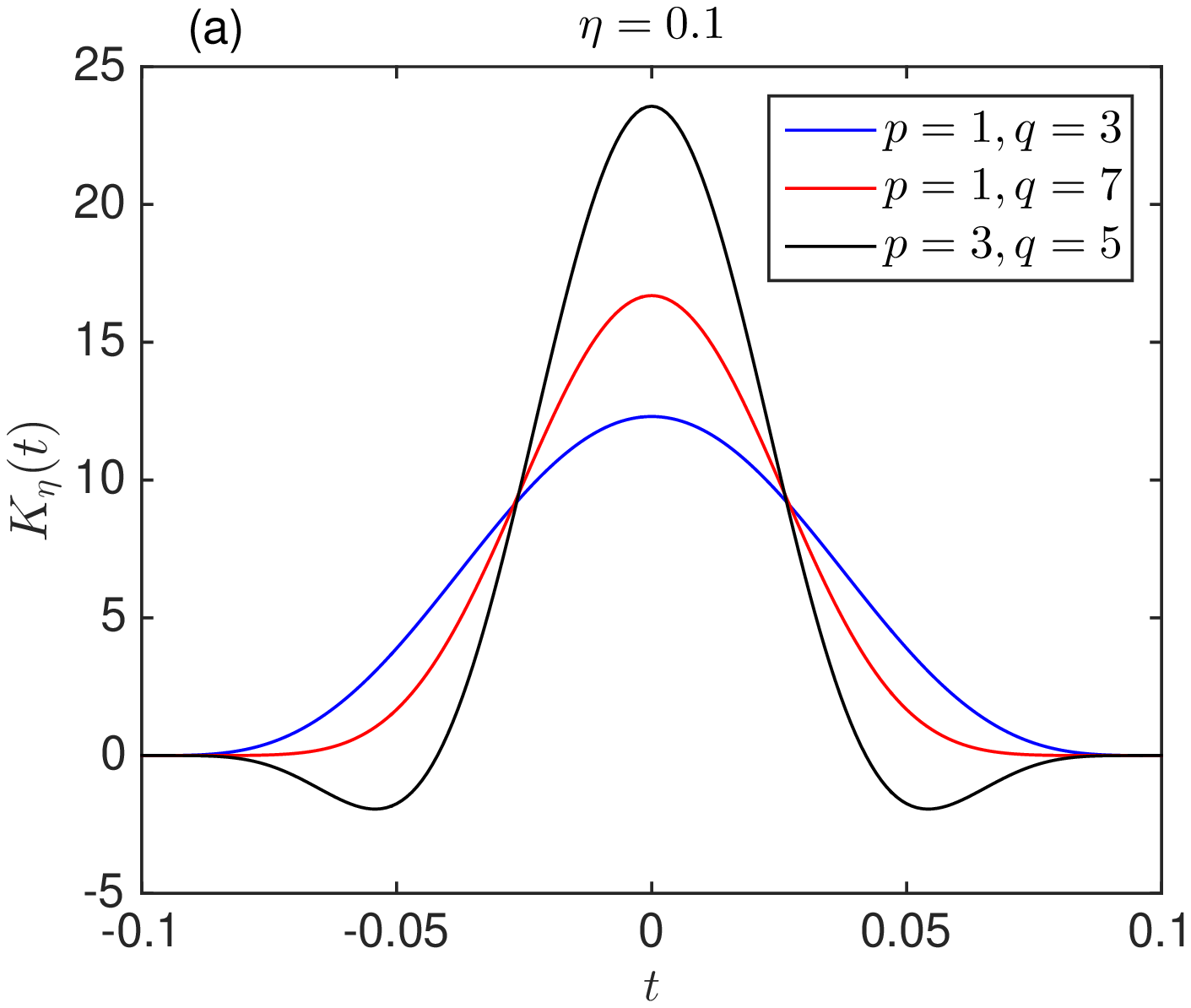}
    \includegraphics[width=0.47\textwidth]{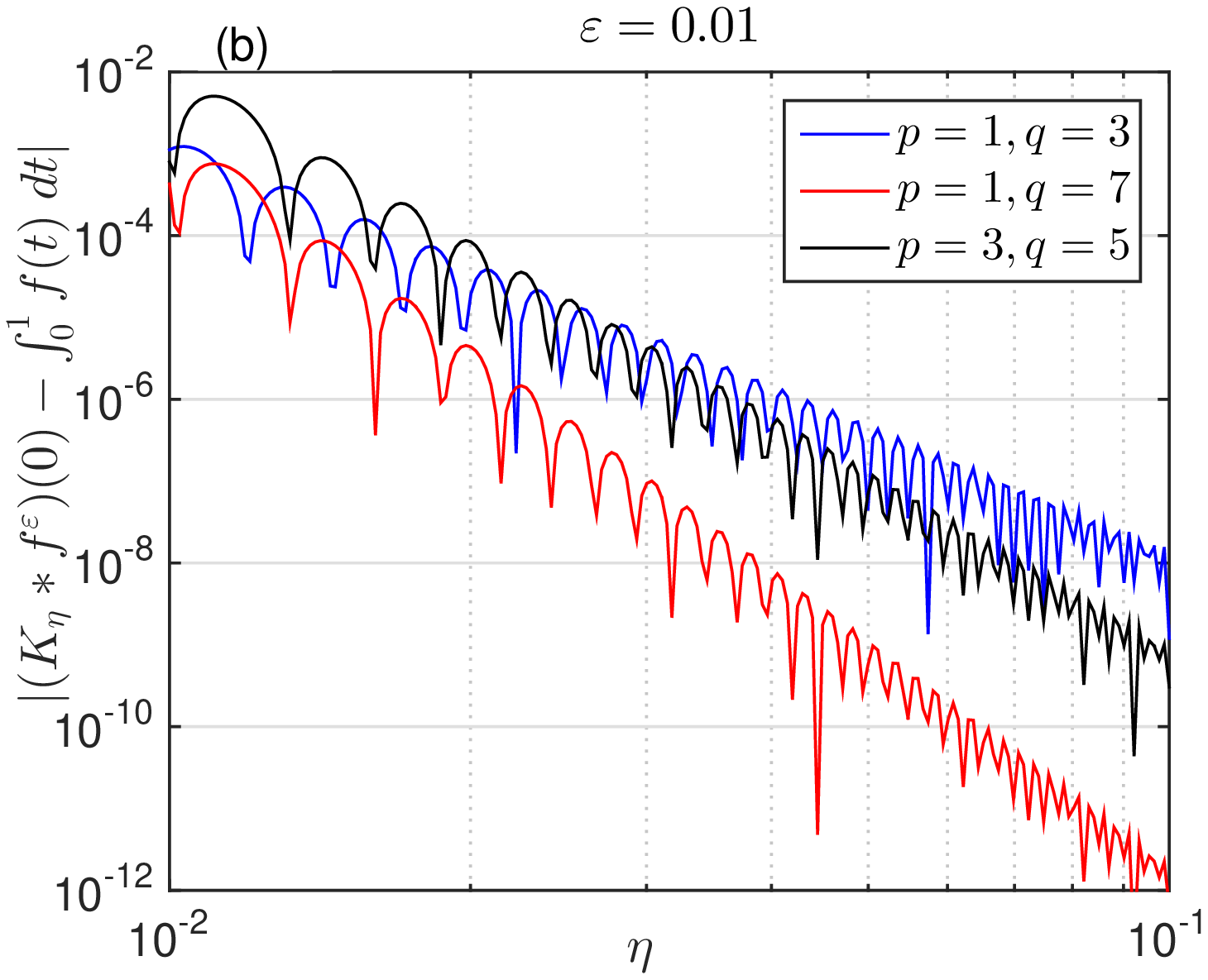}
  \end{center}
  \caption{Averaging kernels $K_{\eta}$ with different values for $p$ and $q$ (a). The corresponding averaging errors for a function $f^\e(t) = \sin(2\pi t/\e)^2$ with $\e = 0.01$ (b).}
  \label{fig:Kernels}
\end{figure}

To simplify the notation, a scaled version of the kernel, denoted by $K_{\eta}(t):=\eta^{-1} K(t/\eta)$, and the shorthand notation 
\begin{equation*}
  \left( K_{\eta} \ast f\right) (t^{\star}) := \int_{t^{\star}-\eta/2}^{t^{\star}+\eta/2} K_{\eta}(t- t^{\star}) f(t) \; \mathrm{d}t
\end{equation*}
are also introduced. Using a kernel $K \in \mathbb{K}^{p,q}$, it is possible to achieve higher order convergence rates and improve the first order approximation \eqref{eq_NaiveAveraging_Estimate}, see e.g. \cite{Arjmand_Runborg2014,Arjmand_Runborg2016a} for theoretical results where the following inequality can also be found:
\begin{equation*}
  \left| \int_{-\eta/2}^{\eta/2} K_{\eta}(t) f(t/\e) \; \mathrm{d}t  - \int_{0}^{1} f(z) \; \mathrm{d}z \right| \leq C \left( \dfrac{\e}{\eta} \right)^{q+2}.
\end{equation*}
In Figure \ref{fig:Kernels}, the decay of the error for kernels with different $p,q$ pairings is shown. In the remaining part of this section, the averaging kernels, which are described above, are used to design an upscaling-based strategy for an atomistic-continuum coupling. 

\subsubsection{HMM}

The problem setting is as follows: the atomistic spins interact with each other with a certain exchange parameter $J_{ij}$ and each spin magnetic moment is subjected to an external field $\vectorn{H}_{{\mathrm e},i}^{\e}(t)  = \vectorn{H}_{{\mathrm e},i}(t,t/\e)$, varying at a slow and a fast scale, where $\e$ represents the size of the fast variations. In this setting, a time step small enough to resolve the $\e$-scale variations of the external field is needed to perform an atomistic spin dynamics simulation. Moreover, to compute a statistical average of the particle magnetisation, a large number of samples of an already expensive atomistic problem have to be computed. Note that the above structure for external field $\vectorn{H}_{{\mathrm e},i}^{\e}(t)$ is chosen only as an example, and the same methodology would work equally well for other types of external fields, which have a scale separation of some form, with no conceptual change in the algorithm.

On the other hand, an upscaling-based multiscale method aims at reducing the degrees of freedom, in time and space, by introducing a discretised macroscopic magnetisation field which is related to the local spatial and temporal averages of the spin magnetic moments. From a modelling point of view, one additional requirement is to include the effect of the thermal noise on the magnetisation. 

\begin{figure}
  \begin{center}
    \includegraphics[scale=0.55]{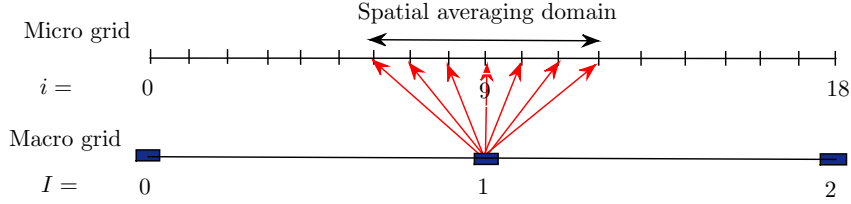}
  \end{center}
  \caption{An example of a macroscopic grid corresponding to a microscopic grid with $r = 3$ and $\ell = 6$. The magnetisation at each macroscopic point (solid blue nodes) is found by averaging the spin magnetic moments around a local spatial averaging domain and a local in time averaging interval (temporal averaging is not shown in the figure).}
  \label{fig_MicroMacroGrid}
\end{figure}

To define the variables, the atomistic spin magnetic moments $\vectorn{m}^{\e}_i$ are introduced. These magnetic moments are located on $\left((r+\ell)N+1\right)$ discrete points defined by $\{ x_{i} = i a \}_{i=0}^{(r+\ell)N}$, where $a$ is the distance between the particles, $r,N \in \mathbb{Z}^{+}$ are positive integers and $\ell$ is a non-negative integer. The superscript $\e$ on $\vectorn{m}^{\e}_i$ indicates the dependence of the spin magnetic moments on the high-frequency variations in the external field $\vectorn{H}^{\e}_{{\mathrm e},i}$. Furthermore, a set of $N+1$ macroscopic variables $\vectorn{M}_{I}$ located on the macro grid $\{ X_{I} = I (r+\ell) a \}_{I=0}^{N}$ are introduced. In Figure \ref{fig_MicroMacroGrid}, an example of coupled grids with $r=3$ and $\ell=6$ is shown. The local mean magnetisations are defined as 
\begin{align}
  \label{eq:LocalAverage_Magnetisation}
  \tilde{\vectorn{M}}_{I}(t)  &=  \sum_{j=-r}^{r} K_{\eta}\left( x_{I(r+\ell) + j}  - x_{I(r+\ell)} \right) \left( K_{\tau} \ast  \vectorn{m}^{\e}_{I(r+\ell)+j} \right)(t) \nonumber \\&=: \left( \mathcal{K}_{\eta,\tau} \ast \vectorn{m}^{\e} \right)(X_{I},t), \quad \text{ where } \eta = 2 r a,  \text{ and } \tau > \e. 
\end{align}
For simplicity, it is assumed that the atomistic particles are globally constrained by periodic boundary conditions, which makes the above definition of $\tilde{\vectorn{M}}_{I}(t)$ well-defined also for the endpoints; namely $I=0$ and $I=N$. The local means $\tilde{\vectorn{M}}_{I}$ are formed by taking the local averages (not only in time but also in space) of $2 r  +1$ particles. To compute $\vectorn{M}_{I}$, a number of realisations of $\tilde{\vectorn{M}}_{I}$ are needed, which results in a reduction of the macroscopic magnetisation length. 

The key idea behind the HMM algorithm for finite temperatures is separate calculation of the directions and the lengths of the expected values of spin magnetic moments, which are ${\hat{\vectorn{M}}}_{I}$ and $s(t)$, respectively. The macroscopic quantities are introduced in the following way: $\vectorn{M}_{I}:= s(t) {\hat{\vectorn{M}}}_{I}$, where ${\hat{\vectorn{M}}}_{I}$ is equal to the local average \eqref{eq:LocalAverage_Magnetisation} but computed at zero temperature\footnote{Although ${\hat{\vectorn{M}}}_{I}$ is computed at zero temperature, the directions of spins at macroscopic nodes are still influenced by the temperature in a nonlinear way via boundary conditions for the microscopic problem. This is evident from the finite-temperature HMM algorithm to be introduced shortly.}, and $s(t)$ is a variable which describes the effect of the temperature in the system.

An algorithm for computing ${\hat{\vectorn{M}}}_{I}$ for a single spin at zero temperature (namely when a single spin is subjected to a high-frequency field and the local averaging takes place only in time) was proposed in \cite{Arjmand2016}. At zero temperature, $s(t) = 1$ holds, and therefore $\vectorn{M} = \hat{\vectorn{M}}$, and the algorithm from \cite{Arjmand2016} is presented below (index $I$ is omitted here).
\bigskip

\noindent {\emph{{\bf Algorithm 1:} A single spin subjected to a deterministic oscillatory field.}}

{\bf Step 1.} \emph{Macro problem: the macroscopic model is incomplete and is given by}
\begin{eqnarray*}
  \label{eq:Macromodel_SingleSpin}
  \dfrac{\mathrm{d}}{\mathrm{d}t} \vectorn{M}(t) = -\vectorn{F}(t,\vectorn{M}) - \dfrac{\alpha_\mathrm{L}}{\beta_\mathrm{L}} \vectorn{M} \times \vectorn{F}(t,\vectorn{M}), \quad \vectorn{M}(0) = \vectorn{m}^{\e}(0),\nonumber
\end{eqnarray*}
\emph{where $\vectorn{F}$ is the missing data in the model and $\vectorn{M}$ is the macro solution to be computed.} 
\bigskip

{\bf Step 2.} \emph{Micro problem: to compute $\vectorn{F}(t_\mathrm{a},\vectorn{M})$ and close the macro problem, the micro problem is solved first}
\begin{eqnarray*}
  \label{eq:Micromodel_SingleSpin}
 \dfrac{\mathrm{d}}{\mathrm{d}t} \vectorn{m}^{\e}(t) =  \beta_\mathrm{L} \vectorn{m}^{\e} \times  \vectorn{H}^{\e}(t), \quad t \in I_{\tau}^{\pm}, \quad \vectorn{m}^{\e}(t_\mathrm{a}) = \vectorn{M}, \nonumber 
\end{eqnarray*}
\emph{where $I^{+}_{\tau}:=(t_\mathrm{a},t_\mathrm{a} + \tau/2]$ and $I^{-}_{\tau}:= [t_\mathrm{a}-\tau/2,t_\mathrm{a})$, and $\tau > \e$. }
\bigskip

{\bf Step 3.} \emph{Upscaling: the unknown parameter $\vectorn{F}(t_\mathrm{a},\vectorn{M})$ in the macro model is computed by}
\begin{eqnarray*}
  \label{eq:UpscalingSingleSpin}
  \vectorn{F}(t_\mathrm{a},\vectorn{M})  &=& \left(  K_{\tau} \ast \dfrac{\mathrm{d}}{\mathrm{d}t} \vectorn{m}^{\e} \right) (t_\mathrm{a}) = \int_{t_\mathrm{a}-\tau/2}^{t_\mathrm{a} + \tau/2} K_{\tau}(t - t_\mathrm{a}) \dfrac{\mathrm{d}}{\mathrm{d}t} \vectorn{m}^{\e}(t)  \; \mathrm{d}t. \nonumber
\end{eqnarray*}
\bigskip

\begin{Remark}
\label{Rem_MagLength}
Starting from a spin magnetic moment $\vectorn{m}^{\e}(0)$ with unit length, it was shown in \cite{Arjmand2016}, that the length of the macroscopic magnetisation $\vectorn{M}$ satisfies
\begin{equation*}
  \left| \vectorn{M}(t) \right| = 1 + O\left( \left( \dfrac{\e}{\tau} \right)^{q+2} \right).
\end{equation*}
Namely, the length of the macroscopic magnetisation is equal to one, up to an upscaling error.
\end{Remark}

\begin{Remark}
\label{Rem:Cost_Alg1}
In \cite{Arjmand2016}, it is shown that when $\vectorn{H}^{\e}(t):= \vectorn{H}(t,t/\e)$, where $\vectorn{H}(t,\cdot)$ is a $1$-periodic smooth function, the error between the HMM (algorithm 1) and the exact macroscopic dynamics is given by
\begin{equation*}
  \text{\textup{Error}}_{\text{\textup{HMM}}} \approx C \left( \underbrace{\dlt t^{r_1}}_{\text{\textup{Macro error}}} + \underbrace{ \left( \tau + \left( \dfrac{\e}{\tau} \right)^{q+2} \right)}_{\text{\textup{Upscaling error}}} + \underbrace{ \left( \dfrac{\delta t}{ \e } \right)^{r_2}}_{\text{\textup{Micro error}}} \right), 
\end{equation*}
where $r_1$ and $r_2$ represent the order of accuracies for a macroscopic and a microscopic solver. Here, $\dlt t$ and $\delta t$ are the macroscopic and the microscopic step sizes. The total computational cost (of HMM) to simulate the dynamics until a time $T$ is given by $T \tau / \left( \dlt t \delta t \right)$. Let $\beta:= 1/ \left( q+3 \right)$. Then the computational cost to achieve an error tolerance $\text{\textup{TOL}} = O(\e^{1-\beta})$ becomes
\begin{equation*}
  \text{\textup{Cost}}_{\text{\textup{HMM}}} \approx T \e^{-(1-\beta)(\frac{1}{r_1}  + \frac{1}{r_2})} \e^{-\beta}.
\end{equation*}
On the other hand, a direct numerical simulation (DNS) gives an error $\left( \delta t / \e \right)^{r_2}$ (with a computational cost $T/\delta t$). Therefore, the cost to achieve the same error tolerance becomes
\begin{equation*}
  \text{\textup{Cost}}_{\text{\textup{DNS}}} \approx T \e^{-\frac{r_2 + 1 - \beta}{r_2}}.
\end{equation*}
The parameter $\beta$ can be made small by taking large values for $q$, which does not increase the computational cost of HMM. Moreover, It is clear from the above estimates that the computational cost of the HMM decreases when higher order methods are used for the macro and the micro solvers, while cost of the DNS can not get better than $O(\e^{-1})$.
\end{Remark}

In algorithm 1 above, the damping term appears only in the macro model (but not the micro problem). However, it was shown (numerically and analytically) that the HMM still captures the correct macroscopic dynamics and hence the correct damping effects, see \cite{Arjmand2016} for the numerical analysis and a motivation for the algorithm. 

The aim here is to present a generalisation of algorithm $1$ to a system of particles interacting at non-zero temperature. To describe the multiscale algorithm, the notations
\begin{align}
  \label{eq:DeterministicOscillatoryH}
  \vectorn{H}_{\mathrm{det},i}^{\e} = \dfrac{1}{\mu} \left(  \sum_{j} J_{ij} \vectorn{m}^{\e}_{j} \right) + \dfrac{1}{\mu} K_{\mathrm{a}} \vectorn{p}_\mathrm{a} \vectorn{p}_\mathrm{a} \cdot \vectorn{m}_i^{\e}+ \vectorn{H}_{\mathrm{e},i}^{\e}(t),
\end{align}
and
\begin{align}
  \label{eq:StochasticH}
  \vectorn{H}_{\mathrm{sto},i}(t,t^{\ast})= \dfrac{1}{\mu} \left(  \sum_{j} J_{ij} \vectorn{m}_{j} \right) + \dfrac{1}{\mu} K_{\mathrm{a}} \vectorn{p}_\mathrm{a} \vectorn{p}_\mathrm{a} \cdot \vectorn{m}_i+ \left( K_{\tau} \ast\vectorn{H}_{e,i}^{\e} \right)(t^{\ast}) + \vectorn{h}_i(t)
\end{align}
are adopted. Note the differences between  $\vectorn{H}_{\mathrm{det},i}^{\e}$ and $\vectorn{H}_{\mathrm{sto},i}$. The term $\vectorn{H}_{\mathrm{det},i}^{\e}$ is deterministic but oscillatory, while $\vectorn{H}_{\mathrm{sto},i}$ is stochastic  and includes the filtered external field $\left( K_{\tau} \ast\vectorn{H}_{\mathrm{e},i}^{\e} \right)(t^{\ast}) $. The notations used for the terms, other than $\vectorn{H}^{\e}_{\mathrm{e},i}$, in \eqref{eq:DeterministicOscillatoryH} and \eqref{eq:StochasticH} are the same as those in \eqref{eq:ASD_LLG}, \eqref{eq:ASD_param} and \eqref{eq:ASD_H}.

\begin{figure}
  \begin{center}
    \includegraphics[scale=0.5]{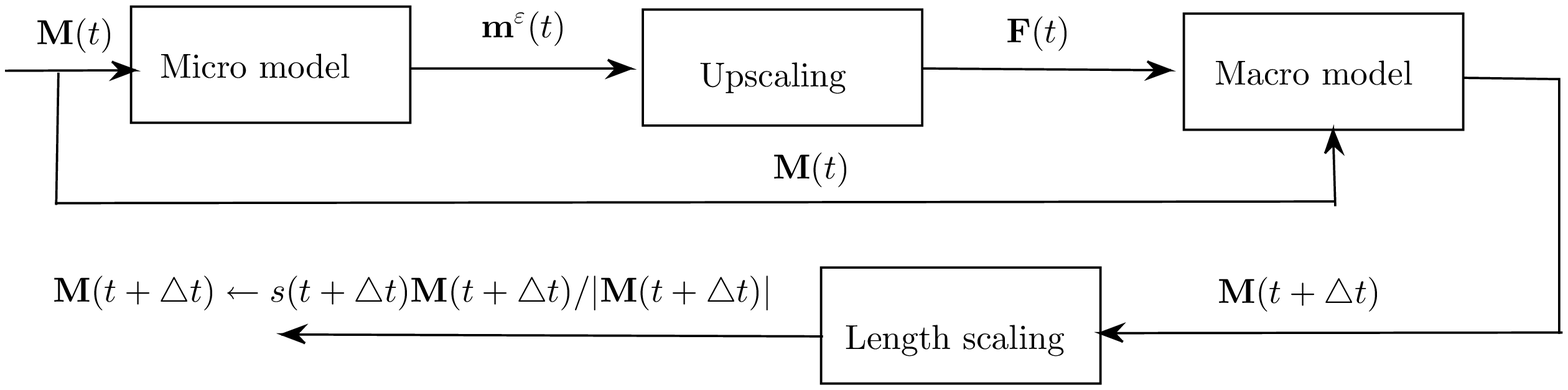}
  \end{center}
  \caption{The input-output relation for the multiscale algorithm to approximate the macroscopic dynamics at finite temperature. One step of the method (assuming a suitable time-stepping) is depicted. If the macro solver is explicit, the steps are followed in the precise order starting from a given $\vectorn{M}(t)$. If the macro solver is implicit, the steps  are repeated until the solution ${\vectorn M}(t + \dlt t)$ (before the length scaling step) converges. Once the convergence is achieved, the length scaling procedure takes place.}
  \label{fig_SchematicAlgorithm}
\end{figure}

To model the macroscopic magnetisation dynamics of a system of magnetic particles at finite temperature, a macro model, a micro model, an upscaling step and a length scaling procedure are required (instead of solving the continuum equation directly). In Figure \ref{fig_SchematicAlgorithm}, a schematic description of the method is illustrated. The entire multiscale algorithm is presented below.
\bigskip

\noindent {\emph{{\bf Algorithm 2:} A chain of magnetic particles at nonzero temperature.}}

{\bf Step 1.} \emph{Macro problem: the macroscopic model is deterministic and given by}
\begin{align}
  \label{eq:Macromodel}
  \dfrac{\mathrm{d}}{\mathrm{d}t} \vectorn{M}_{I}(t)  & =   -\vectorn{F}_{I}(t,\vectorn{M}_{I^\prime}) -\dfrac{\alpha_\mathrm{L}}{\beta_\mathrm{L}} \vectorn{M}_{I} \times \vectorn{F}_{I}(t, \vectorn{M}_{I^\prime}), \nonumber  \\ \vectorn{M}_I(0) &= s_{I}(0) \vectorn{m}^{\e}_{Ir^{\prime}}(0),
\end{align}
\emph{where $r^{\prime} = r + \ell$, and $\vectorn{F}_{I}$ is the missing data in the model and $\vectorn{M}_I$ is the  macro solution to be computed. Moreover, $I^{\prime}:=\left\{ I-1,I,I+1\right\}$ represents the index set of the nearest macro solutions. The computation of $s_I(t)$ is described in Step $4$ below. } 
\bigskip

{\bf Step 2.} \emph{Micro problem:  to compute $\vectorn{F}_I(t_{\mathrm{a}},\vectorn{M})$ and close the macro problem, the micro problem is solved at first}
\begin{align}
  \label{eq_Micromodel}
  \dfrac{\mathrm{d}}{\mathrm{d}t} \vectorn{m}^{\e}_{I r^{\prime}+j}(t)  &= \beta_\mathrm{L} \vectorn{m}^{\e}_{Ir^{\prime}+j}(t) \times \vectorn{H}^{\e}_{\mathrm{det},Ir^{\prime}+j}(t), \quad t \in I^{\pm}_{\tau}, \quad  j =-r,\ldots,r, \nonumber \\
  \vectorn{m}^{\e}_{Ir^{\prime}+j}(t_\mathrm{a}) &= \hat{\vectorn{m}}(x_{Ir^{\prime}+j}), \\
  \vectorn{m}^{\e}_{Ir^{\prime} - r}(t) &=  \hat{\vectorn{m}}(x_{Ir^{\prime} - r}), \quad
  \vectorn{m}^{\e}_{Ir^{\prime}+r}(t) = \hat{\vectorn{m}}(x_{Ir^{\prime} + r}), \nonumber
\end{align}
\emph{where $I^{+}_{\tau}:=(t_\mathrm{a},t_\mathrm{a}+\tau/2]$ and $I^{-}_{\tau}:= [t_\mathrm{a}-\tau/2,t_\mathrm{a})$,  with $\tau > \e$,  and $\hat{\vectorn{m}}(x) = \pi_2 \vectorn{M}/|\pi_2 \vectorn{M}|(x)$ denotes the normalised second order polynomial interpolation of the macroscopic solutions.}
\bigskip

{\bf Step 3.} \emph{Upscaling: the parameter $\vectorn{F}_I(t_\mathrm{a},\vectorn{M}_{I^{\prime}})$ is computed by}
\begin{align}
  \label{eq:Upscaling}
  \vectorn{F}_{I}(t_\mathrm{a},\vectorn{M}_{I^{\prime}})  =\sum_{j=-r}^{r}   K_{\eta}(x_j-x_{Ir^{\prime}})  \int_{t_\mathrm{a}-\tau/2}^{t_\mathrm{a} + \tau/2} K_{\tau}(z - t_\mathrm{a}) \dfrac{\mathrm{d}}{\mathrm{d}z} \vectorn{m}_{Ir^{\prime}+j}^{\e}(z)  \; \mathrm{d}z, 
\end{align}
where $\eta = 2 r a$.
\bigskip

{\bf Step 4.} \emph{Length scaling: the macro solution $\vectorn{M}_I(t)$ is replaced by $s_I(t) \vectorn{M}_{I}(t)/ |\vectorn{M}_{I}(t)|$, where  $s_{I}(t_\mathrm{a})$ is computed by solving the following stochastic LLG equation for $j = -r,\ldots,r$ and $t \in(t_\mathrm{a},t_\mathrm{a}+\tau_\mathrm{f}]$}
\begin{eqnarray}
  \label{eq:LengthScaling}
 \dfrac{\mathrm{d}}{\mathrm{d}t} \vectorn{m}_{Ir^{\prime}+j}(t)  &=& -\beta_\mathrm{L} \vectorn{m}_{Ir^{\prime}+j}(t) \times \vectorn{H}_{\mathrm{sto},Ir^{\prime}+j}(t;t_\mathrm{a})- \alpha_\mathrm{L} \vectorn{m}_{Ir^{\prime}+j} \times (\vectorn{m}_{Ir^{\prime}+j} \times \vectorn{H}_{\mathrm{sto},Ir^{\prime}+j}(t;t_\mathrm{a})), \nonumber\\ 
  \vectorn{m}_{Ir^{\prime}+j}(t_a) &=& \hat{\vectorn{m}}(x_{Ir^{\prime}+j}),\\
  \vectorn{m}_{Ir^{\prime} - r}(t) &=&  \hat{\vectorn{m}}(x_{Ir^{\prime} - r}), \quad \vectorn{m}_{Ir^{\prime} + r}(t) =  \hat{\vectorn{m}}(x_{Ir^{\prime} + r}), \nonumber
\end{eqnarray}
\emph{where $\tau_\mathrm{f} > \tau_\mathrm{r}$, and $\tau_\mathrm{r}$ is the time it takes to reach the thermal equilibrium. Then, with $\eta = 2 r a$, the following is computed}
\begin{align}
  \label{eq:LengthComput}
  s_{I}(t_\mathrm{a})  = \left|  \sum_{j=-r}^{r} K_{\eta}(x_j-x_{Ir^{\prime}}) \left( \dfrac{1}{\tau_\mathrm{f}-\tau_\mathrm{r}} \int_{\tau_\mathrm{r}}^{\tau_\mathrm{f}} \vectorn{m}_{Ir^{\prime}+j} (t) \;\mathrm{d}t \right) \right|. 
\end{align}
\bigskip

\begin{Remark}
Note that the parameter $s_I(t)$ is allowed to vary in time and space. This approach is more general than the mean-field approximation (MFA) since spatially nonuniform magnetisation lengths are automatically captured here, whereas MFA assumes a uniform length everywhere.  
\end{Remark}

\begin{Remark}
The fact that the HMM algorithm, which is described above, uses local averages (in time and space) of the microscopic solution, makes it possible to upscale the influence of other kind of microscopic variations such as defects (modelled by $J_{ij}=0$ for certain $i,j$ pairs) or fast variations in the material properties (modelled by rapid spatial change of $J_{ij}$ or $\vectorn{p}_{\mathrm{a}}$). The resulting macroscopic averages, however, need to be smooth enough so that the macro discretisation resolves it. If not, the atomistic picture needs to be retained, and the partitioned domain approach would be more appropriate.
\end{Remark}

\begin{Remark}
In equation \eqref{eq:LengthComput}, the usage of kernels $K_{\eta}$ (centred at $x_{Ir^{\prime}}$) for spatial averaging is important. At finite temperatures it holds that $s_{I}(t) <1$. However, since the boundary conditions for ${\bf m}$ in \eqref{eq:LengthScaling} are fixed, the length of the temporal average of the boundary atom is always $1$. The usage of a kernel resolves this issue, as $K_{\eta}$ vanishes towards the boundaries of the micro domain, see e.g. Figure \ref{fig:Kernels}. A more secure way of reducing the boundary error is to use a damping band near the boundary, see the equation \eqref{eq:dampBand_ASD_LLG}, where a damping layer is added to reduce the high frequency reflections from the boundary. However, these damping bands did not seem to be necessary for the simulations presented in the next section.
\end{Remark}

\section{Numerical results and discussions}
\label{sec:methods}

\subsection{Domain partitioning}
\label{sec:res_DP}

\subsubsection{Damping region at the atomistic-continuum boundary}

In \cite{Poluektov2016}, a special damping region for domain partitioning multiscale method was suggested. This damping region reduces reflections of high-frequency waves from the atomistic-continuum interface, thereby reduces the numerical error of the solution within the atomistic region of the mutiscale system. The damping band acts as a low-pass filter.

This concept was previously tested only in deterministic cases \cite{Poluektov2016,Poluektov2016b}. In the case when the atomistic region is stochastic, the main requirement for the damping band is not to affect the thermodynamic characteristics of atoms. In the spin dynamics approach, each atom is coupled to the heat bath, which allows using a rather simple damping for a certain layer of atoms, while preserving correct thermodynamics in neighbouring layers. A numerical experiment designed to demonstrate this is presented below.

The dynamic behaviour of the multiscale system, in which the atomistic region is embedded into the continuum region, is simulated. The details regarding the multiscale coupling method and the damping region at the interface can be found in \cite{Poluektov2016b}. Material parameters ($\mu$, $J$, $K_\mathrm{a}$, $\vectorn{p}_\mathrm{a}$, $\lambda$) that were used in the simulation are summarised in Table \ref{tab:sizeParam}, material 2. Atomistic region was located in the centre of the geometry, the entire length of which was $L$. Spatial step of $\dlt x = 4 a$ was used in the continuum model, where $a = 0.3636\cdot10^{-9}$ is the interatomic spacing, as well as the Neumann boundary conditions. The atomistic model consisted of $N$ atoms, $N_\mathrm{D}$ out of which were numerically damped atoms (see \cite{Poluektov2016b}), i.e. $N_\mathrm{D} / 2$ damped atoms at left and right edges of the atomistic region. Numerical damping parameter was selected to be $g_\mathrm{D} \gamma^{0.5} = 41.53$ and the averaging window width was $s_\mathrm{A} = 3 \dlt x$. The parameters are summarised in Table \ref{tab:multiscaleParam}, set 1. The initial state of the system was uniform alignment of all spins. Since this system had only exchange energy, a discrete set of interatomic-link energies was calculated for each time step: $e_i =  J \left( 1 - \vectorn{m}_i \cdot \vectorn{m}_{i+1} \right)$, where only links between (numerically) non-damped atoms were taken into account, thus resulting in $\left(N-N_\mathrm{D}-1\right)$ values.

\begin{figure}
  \begin{center}
    \includegraphics[scale=0.8]{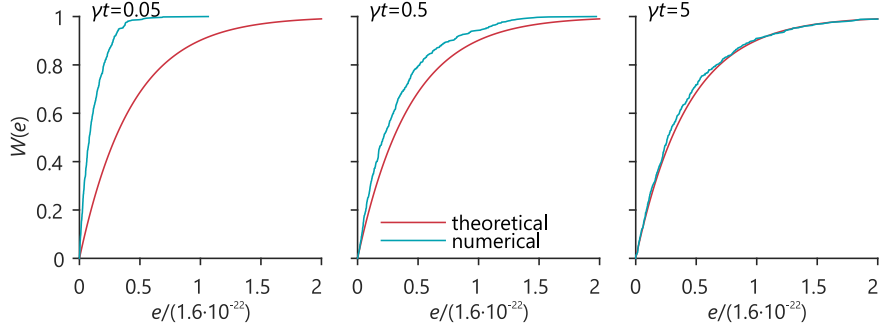}
  \end{center}
  \caption{Cumulative distribution function of interatomic interaction energy calculated at different times. Theoretical line corresponds to the Boltzmann distribution.}
  \label{fig:MSboltz}
\end{figure}

In Figure \ref{fig:MSboltz}, the evolution of cumulative distribution function of interatomic interaction energy ($\varepsilon_i$) is shown.  When system is at the equilibrium, the cumulative distribution of energies should be close to the Boltzmann distribution, $W\left(e\right) = 1 - \exp\left(- e / \left( k_\mathrm{B} T \right) \right)$. As seen from the results, the non-equilibrium initial configuration evolved into equilibrium configuration with the correct energy distribution. This example illustrates that the damping band does not affect the thermodynamic characteristics of the spin system, while reducing wave reflections form the interface. Since it is possible to adjust the parameters of the damping band independently of the geometry and control the error resulting from wave reflections, such damping region can be very useful for coupling dynamic atomistic region to quasistatic continuum region as well.

\begin{table}
  \begin{center}
    \begin{tabular}{l*{8}{r}}
    \hline
    parameter         & $\vectorn{H}_\mathrm{e}$  & $T$ & $N_\mathrm{s}$ & $\dlt t\gamma$ & $\vectorn{m}_i\left(0\right)$ & $L$ & $N$ & $N_\mathrm{D}$ \\
    \hline
    set 1        & $\vectorn{0}$ & $5$ & $1$ & $0.001$ & $\vectorn{e}_y$ & $3840 a$ & $639$ & $32$ \\
    set 2        & $\vectorn{0}$ & $0.1$ & $1000$ & $5$ & \textbf{**} & $256 a$ & $63$ & $32$ \\
    set 3        & \textbf{*} & $30$ & $30$ & $0.01$ & $\vectorn{e}_y$ & $768 a$ & $127$ & $32$ \\
    \hline
    \multicolumn{9}{l}{\textbf{*} - $H_\mathrm{e}\vectorn{e}_z$; $H_\mathrm{e} = 0$ for $t \gamma < 10$; $H_\mathrm{e} = 2$ for $t \gamma \geq 10$} \\
    \multicolumn{9}{l}{\textbf{**} - uniformly changing from $\vectorn{e}_y$ at the left boundary} \\
    \multicolumn{9}{l}{to $\vectorn{e}_z$ at the right boundary}\\
    \end{tabular}
  \end{center}
  \caption{Parameters that were used in the multiscale simulations. Here, $\dlt t \gamma$ is a time step multiplied by the gyromagnetic ratio and $\vectorn{m}_i\left(0\right)$ are the initial directions of $\vectorn{m}_i$.}
  \label{tab:multiscaleParam}
\end{table}

\subsubsection{1D system, switching of the uniform external field}

One of the tests of consistency between atomistic and continuum regions is to check for an identical response of the regions to a non-stationary external field under equilibrium conditions. To illustrate this, a multiscale system, in which the atomistic region is embedded into the continuum region, is  again considered. The system was subjected to an initial applied external field $\vectorn{H}_\mathrm{e} = \vectorn{0}$. When the system reached equilibrium, $t \gamma = 10$, the field was changed to $\vectorn{H}_\mathrm{e} = H_\mathrm{e}\vectorn{e}_z$, $H_\mathrm{e} = 2$, which initiated a precessional motion of the spin magnetic moments. Material parameters ($\mu$, $J$, $K_\mathrm{a}$, $\vectorn{p}_\mathrm{a}$, $\lambda$) that were used in the simulation are summarised in Table \ref{tab:sizeParam}, material 2, while other simulation parameters are summarised in Table \ref{tab:multiscaleParam}, set 3. The Neumann boundary conditions were used in the continuum region. 

\begin{figure}
  \begin{center}
    \includegraphics[scale=0.8]{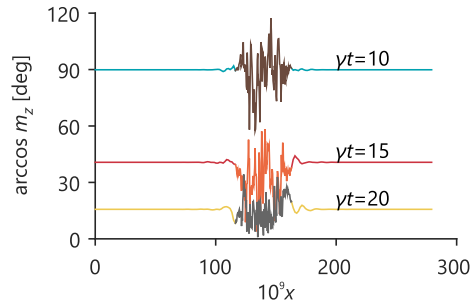}
  \end{center}
  \caption{Ensemble average of the $m_z$ component of the magnetisation as a function of the spatial coordinate at different times. The atomistic region is highlighted with an alternative colour for each curve.}
  \label{fig:MSswitch}
\end{figure}

In Figure \ref{fig:MSswitch}, the behaviour of the magnetisation of the system at thermodynamic equilibrium is shown. It can be seen that the spin moments within the atomistic region rotate with the same angular speed as the continuum region magnetisation, since the average $m_z$ component of the atomistic spins is always close to the continuum value.

Although this is a trivial example, it illustrates that the basic requirement of consistency between models, which are used in the multiscale system, can be fulfilled in certain cases without extension of the continuum description. In this example, parameters of the continuum model corresponded to $0\unit{K}$, and consistency was observed since low temperature was imposed. This is not always the case, as demonstrated in the next example. As was already discussed, the continuum model should be selected depending on the problem to which the multiscale method is applied and sometimes even the basic models provide the solution with an acceptable numerical error. 

\subsubsection{1D system, uniform magnetisation gradient}

The domain partitioning approach also has limitations. These limitations are not related to the coupling between atomistic and continuum regions, but to the nature of the domain partitioning methodology itself. The problem concerns finite-temperature dynamics, where the macroscopic model is dynamic.

Since the atomistic region is stochastic, excitations of different length scale appear in it. Although the probability of large scale excitation, i.e. several spin atomic moments moving in approximately uniform direction, is small, such excitation creates long wave length spin waves, which propagate within the atomistic region. These waves can propagate into the continuum region. In this case, when the ensemble average solution is calculated, a continuum magnetisation with decreased length is obtained. Such long wave length waves cannot be filtered out by the damping band, since it would make modelling dynamics pointless: the single purpose of the damping band is to filter out the waves which cannot be represented by the continuum. In this setting, it is impossible to distinguish between ``useful'' long-range excitations, which have to modelled, and ``unphysical'' ones.

To demonstrate this, a multiscale system, in which the atomistic region is embedded into the continuum region, is considered. Magnetisation at the edges of the continuum domain was fixed, i.e. the stationary Dirichlet boundary conditions were used in the continuum region. Material parameters ($\mu$, $J$, $K_\mathrm{a}$, $\vectorn{p}_\mathrm{a}$, $\lambda$) that were used in the simulation are summarised in Table \ref{tab:sizeParam}, material 2, while other simulation parameters are summarised in Table \ref{tab:multiscaleParam}, set 2. The evolution of the system was simulated until the equilibrium state, which corresponded to $t \gamma = 10^4$, was reached. The final result is obtained by calculating the ensemble average.

\begin{figure}
  \begin{center}
    \includegraphics[scale=0.8]{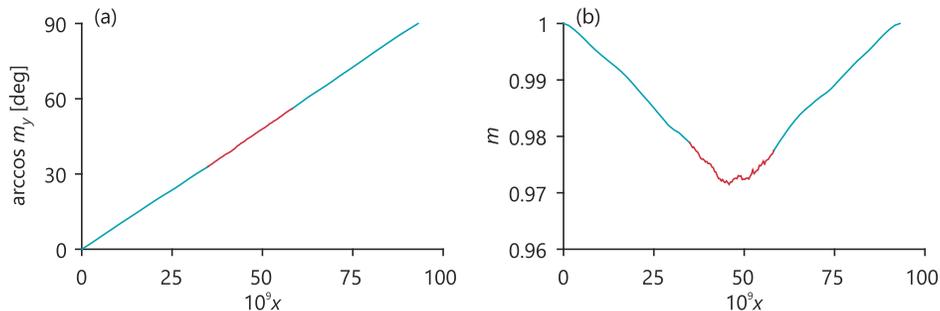}
  \end{center}
  \caption{Ensemble average of $m_y$ component of magnetisation (a) and magnetisation length (b) depending on spatial coordinate. The atomistic region is highlighted with the red colour.}
  \label{fig:MSgrad}
\end{figure}

In Figure \ref{fig:MSgrad}, the state of the system at the equilibrium is shown. It can be seen that the expected value of the magnetisation length\footnote{Within subsection \ref{sec:res_DP}, term ``magnetisation length'' is used for the entire multiscale domain. It is implied that within the atomistic region it means the length of the expected value of the spin magnetic moment.} in the continuum is decreased, see Figure \ref{fig:MSgrad}b, due to waves which are generated within the atomistic region and propagate into the continuum region. The consequences of this issue can be less profound if larger mesh difference is used. Therefore, HMM should not suffer from this issue  due to a relatively large separation of scales. However, this problem can be completely eliminated only by using a quasistatic continuum model. In this case, all waves originating from the atomistic region will be either damped within the damping band region or reflected from the interface.

\subsection{Upscaling via HMM}

In what follows, three different applications of the HMM for non-zero temperatures (Algorithm 2), presented in Section \ref{Upscaling_Subsection}, as well as its limitations are described. \emph{The first example} deals with a single magnetic particle which interacts with a high-frequency external field. As a single particle is studied, i.e. no particle interaction is involved, and the spatial averaging in the algorithm is not used. \emph{The second example} concerns the behaviour of a number of interacting particles. In this example, the number of atomistic particles is chosen equal to the number of macroscopic nodes. The microscopic model \eqref{eq_Micromodel} as well as the stochastic LLG equation \eqref{eq:LengthScaling} are not solved locally (as described above) but globally over the entire domain, and the upscaling step \eqref{eq:Upscaling} is performed only in time. \emph{The third example} corresponds to the case of interacting particles and disparate spatial scales for micro and macro models. Hence all parts of the algorithm (as described in the Section \ref{Upscaling_Subsection}) are utilised, i.e. multiscaling both in time and space is considered. In the last part, the limitations of the HMM are illustrated.

\subsubsection{HMM: Single spin}

\begin{figure}
  \begin{center}
    \includegraphics[width=0.47\textwidth]{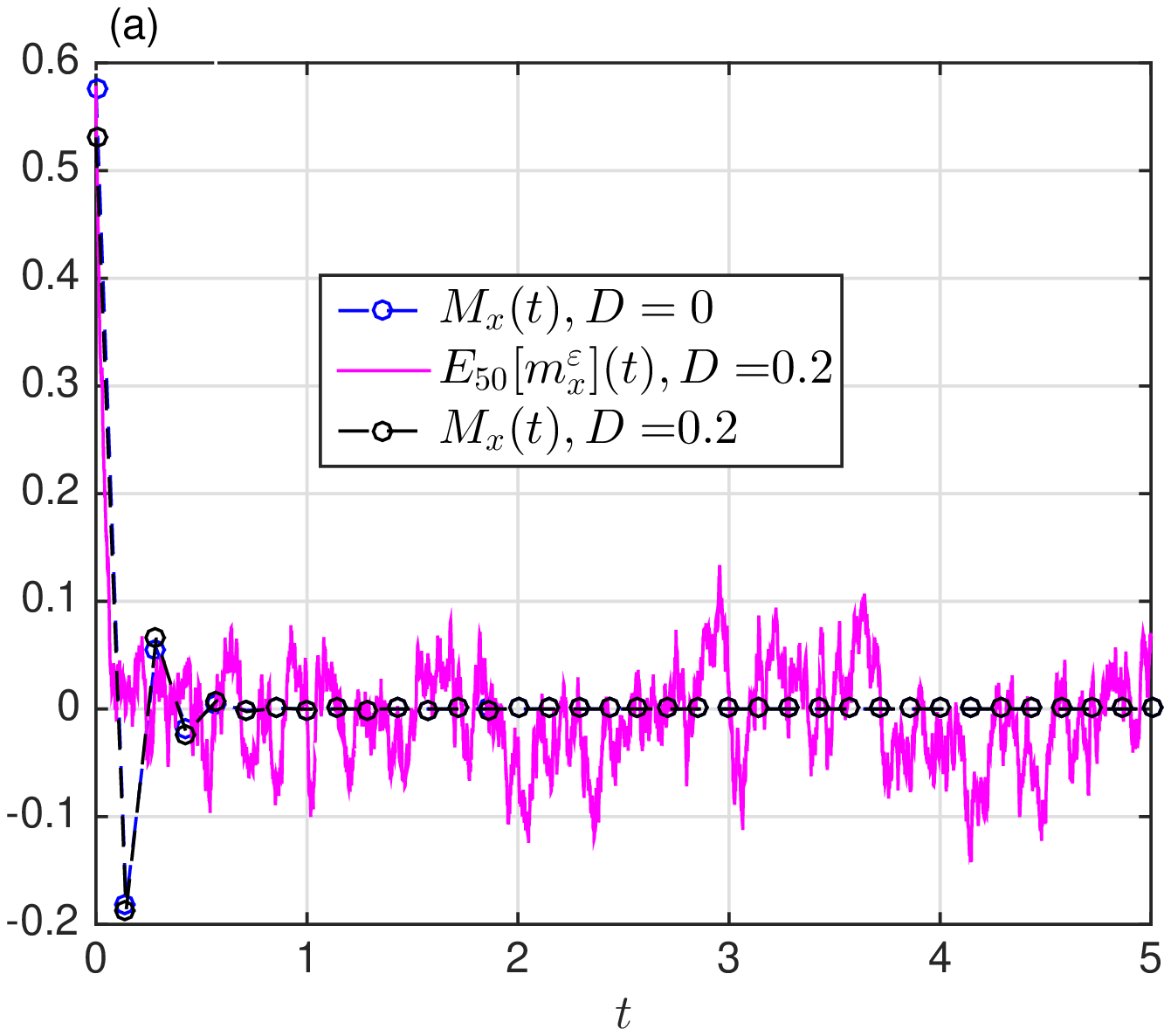}
    \includegraphics[width=0.47\textwidth]{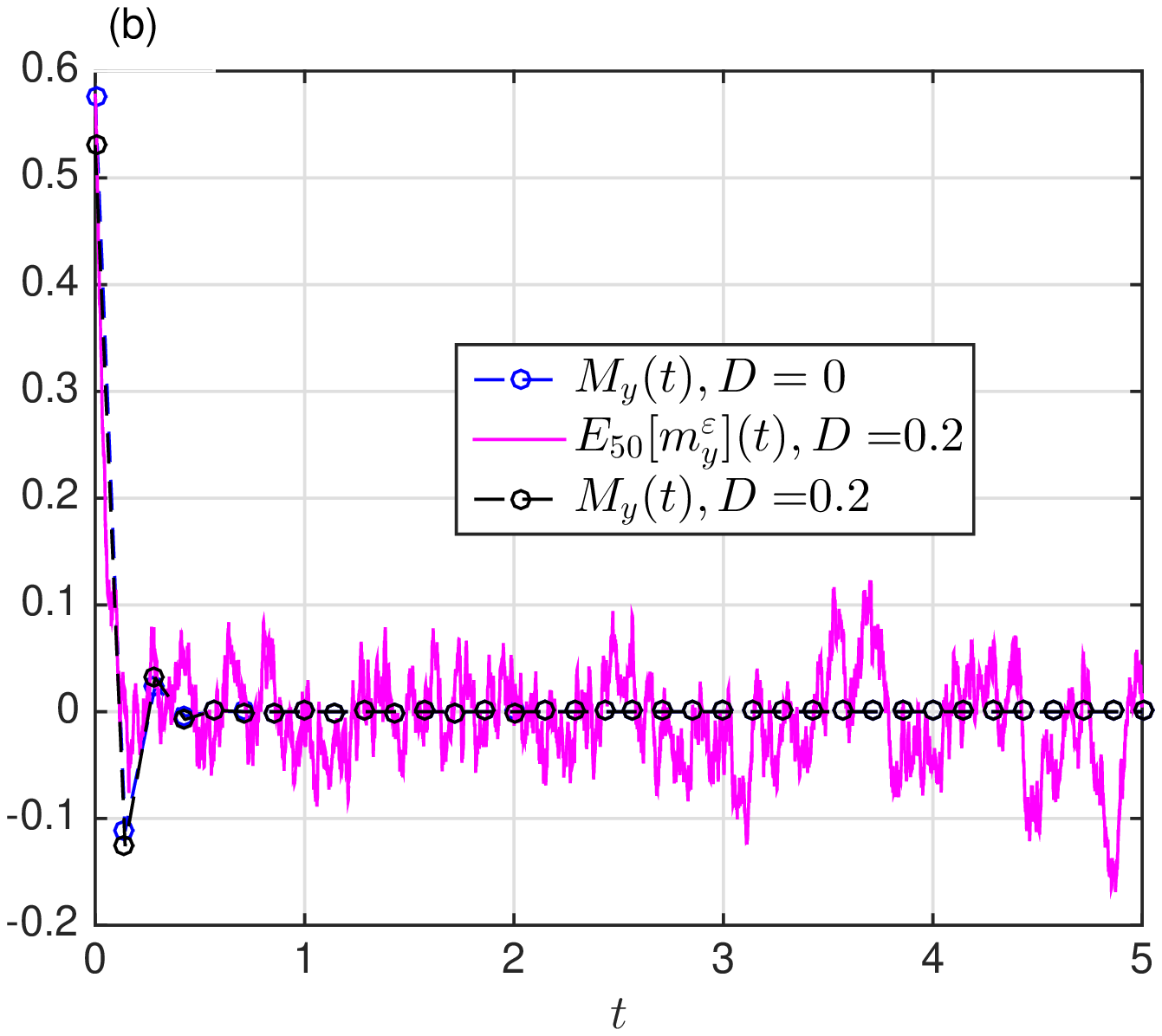}
    \includegraphics[width=0.47\textwidth]{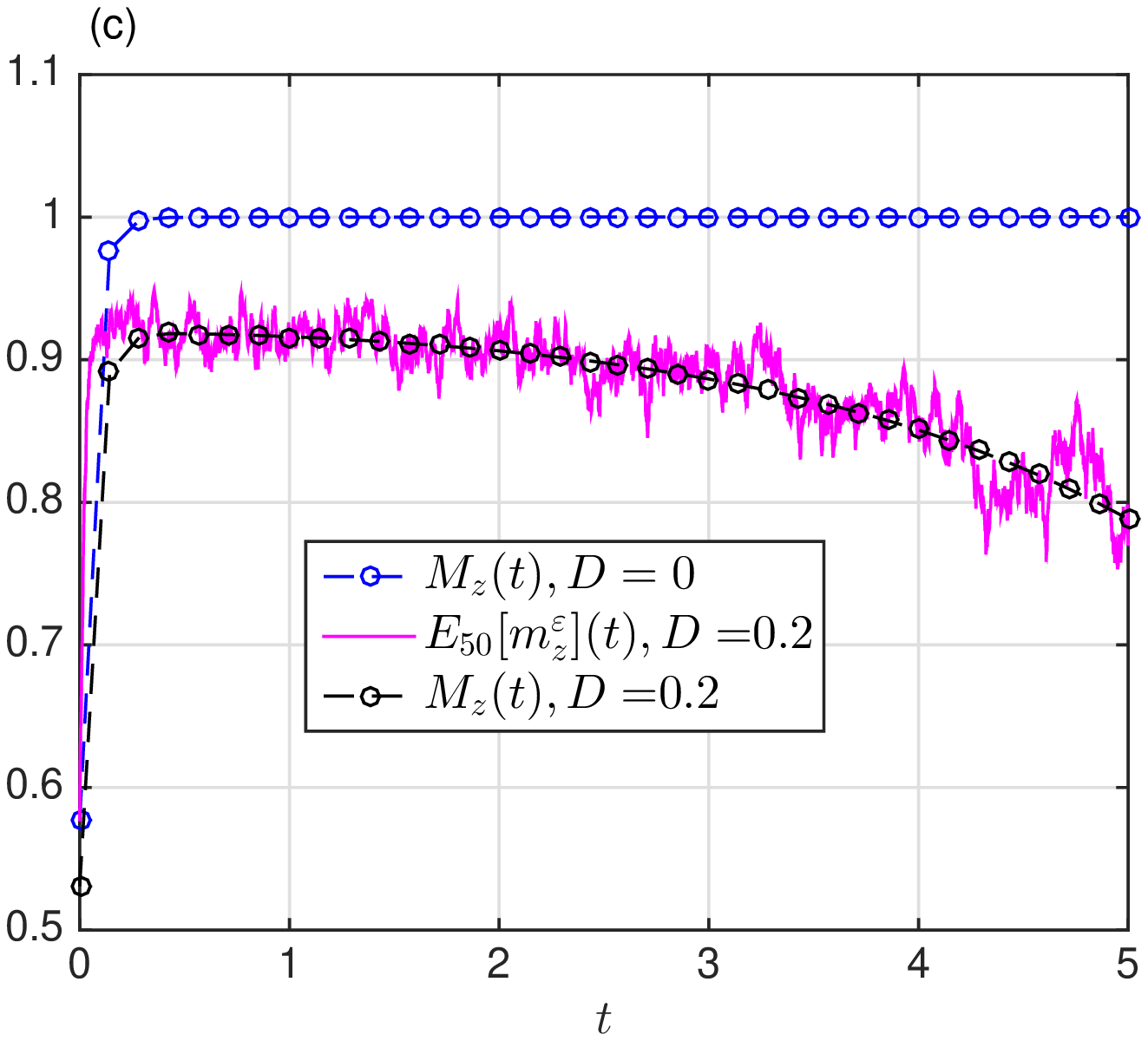}
    \includegraphics[width=0.47\textwidth]{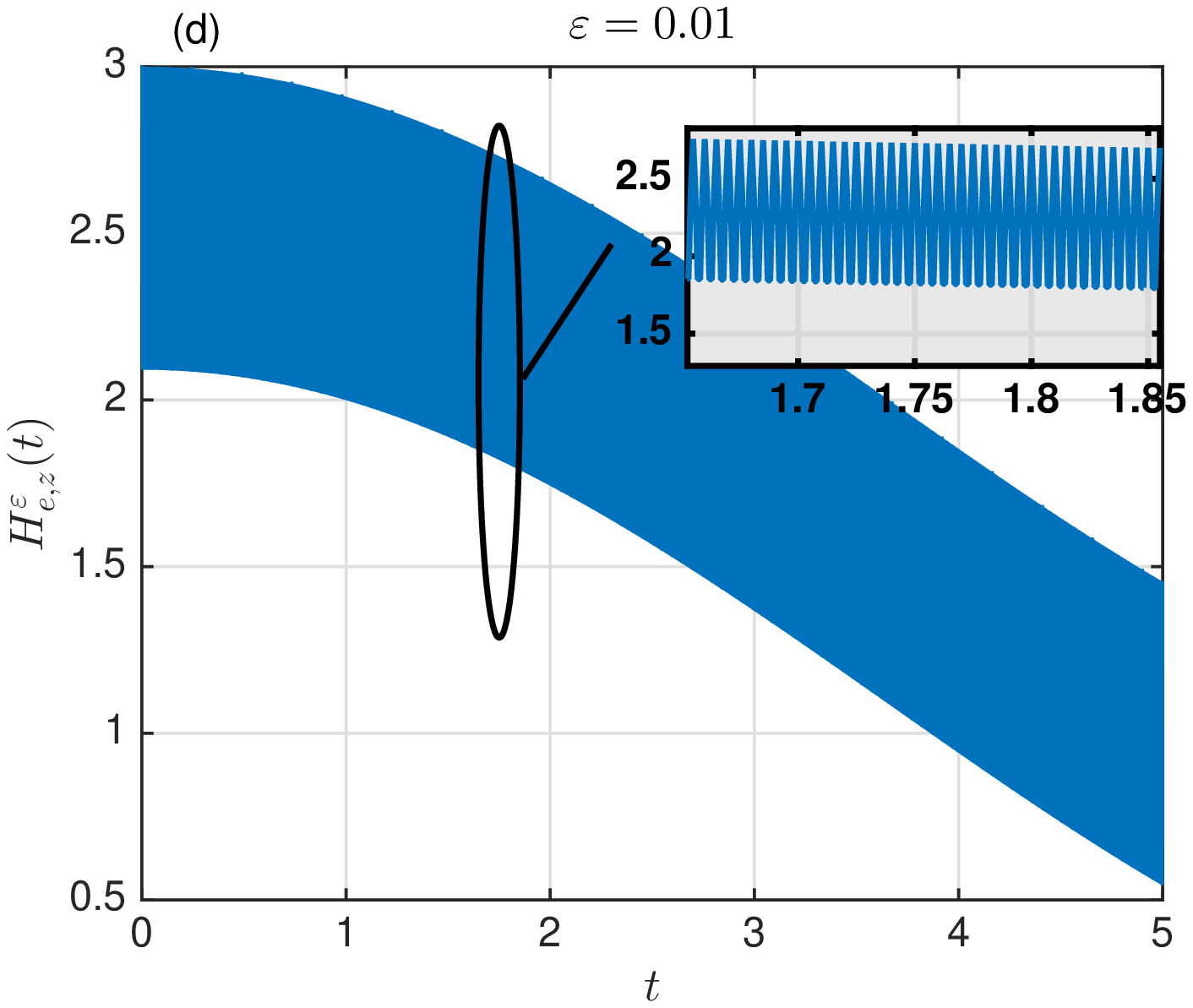}
  \end{center}
  \caption{The components of the spin magnetic moment in directions $x$, $y$, $z$ (a-c): HMM simulation for the case of $D=0$ and $D=0.2$; the exact temperature dynamics computed by averaging $50$ replicas of the stochastic LLG equation \eqref{eq:ASD_LLG} with the external field $\vectorn{H}_{\mathrm{e}}^{\e}$ given by \eqref{eq:ExternalField}. The value $\e = 0.01$ is used in these simulations. The $z$-component of the external field in \eqref{eq:ExternalField} (d). }
  \label{fig:SingleSpin}
\end{figure}

In this first example, a single magnetic particle is subjected to a high frequency field which is nonzero in the $z$-direction. Namely, 
\begin{equation}
  \label{eq:ExternalField}
  \vectorn{H}_{\mathrm{e}}^{\e} = (1+\cos(0.43  t)+\cos(2\pi t/\e)^2)\vectorn{e}_z ,
\end{equation}
while the parameters in \eqref{eq:ASD_LLG} are set to $K_\mathrm{a} = J_{ij} = 0$. Initially, the direction of the spin magnetic moment is set to $\frac{1}{\sqrt{3}} (\vectorn{e}_x+\vectorn{e}_y+\vectorn{e}_z)$. When thermal equilibrium is reached, the solution has to be in the direction of the applied field ($z$-direction). The external field is varying rapidly over time, and hence it influences the evolution of the length of magnetic moment\footnote{Here, as in various other parts of this paper, the expected values of spin magnetic moment is implied.}. In Figure \ref{fig:SingleSpin}, it can be seen that the HMM captures the evolution computed by a direct numerical simulation of the spin length. In the same Figure, the HMM solution at zero temperature (with $s(t) = 1$ or $D = 0$) is shown to emphasise that if the length scaling part  \eqref{eq:LengthScaling} of the HMM is not used, then the method does not follow the correct dynamics, see Figure \ref{fig:SingleSpin}c. For these simulations, an implicit mid-point rule with $35$ macroscopic time steps is used but other numerical integration methods may also be used without any conceptual change in the algorithm, see e.g. \cite{Cervera2007,dAquino2005,Cimrak2008} for a review of numerical methods in micromagnetism. The macroscopic magnetisation is evolved until $t = 5$, and the micro problems are solved over an interval of size $\tau = 5 \e$. The physical parameters are chosen to be $\beta_\mathrm{L}= 1$, $\alpha_\mathrm{L} = 10$ and $D = 0.2$. 

\subsubsection{HMM: Chain of spins, multiscaling only in time}

\begin{figure}
  \begin{center}
    \includegraphics[width=0.47\textwidth]{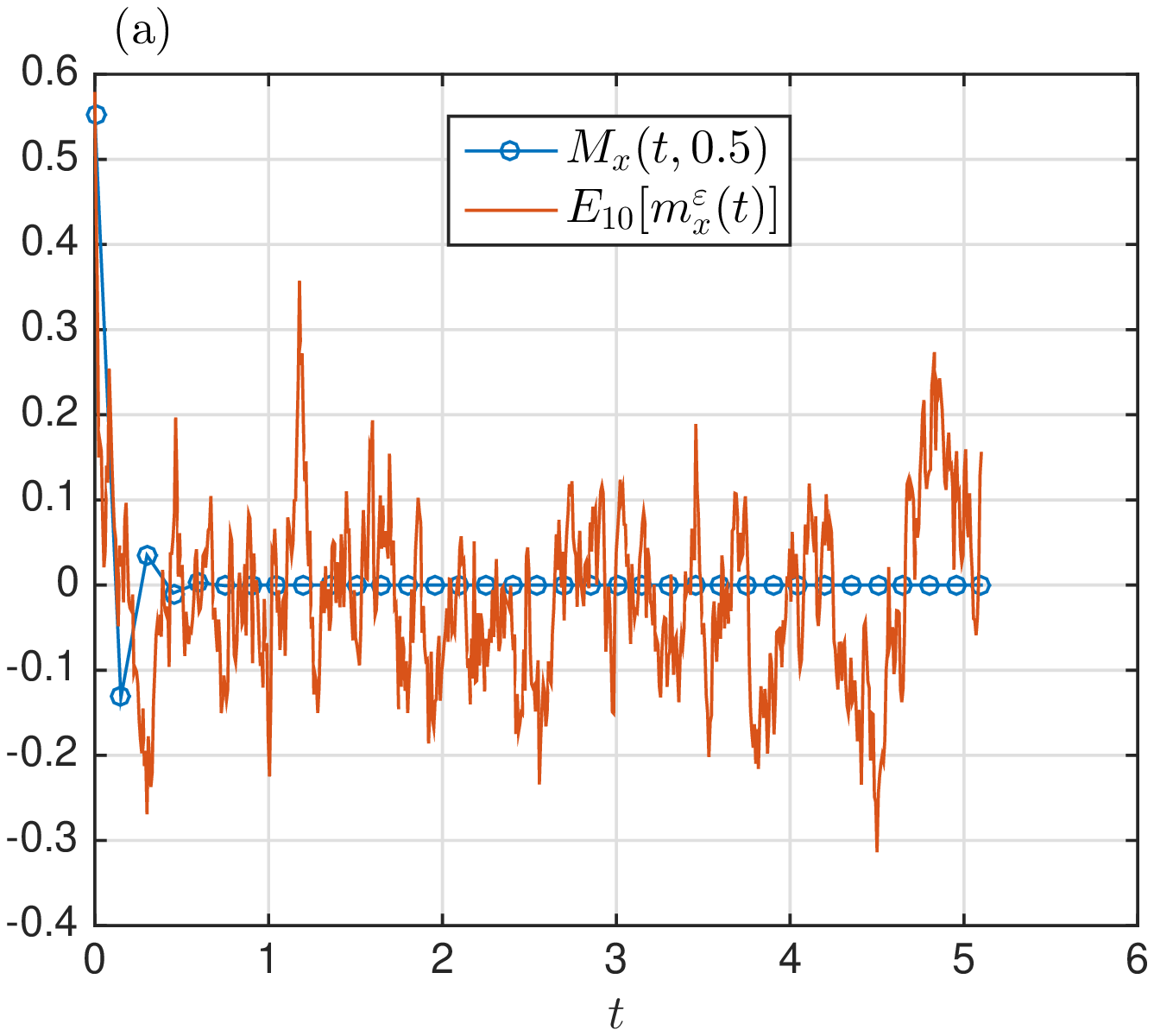}
    \includegraphics[width=0.47\textwidth]{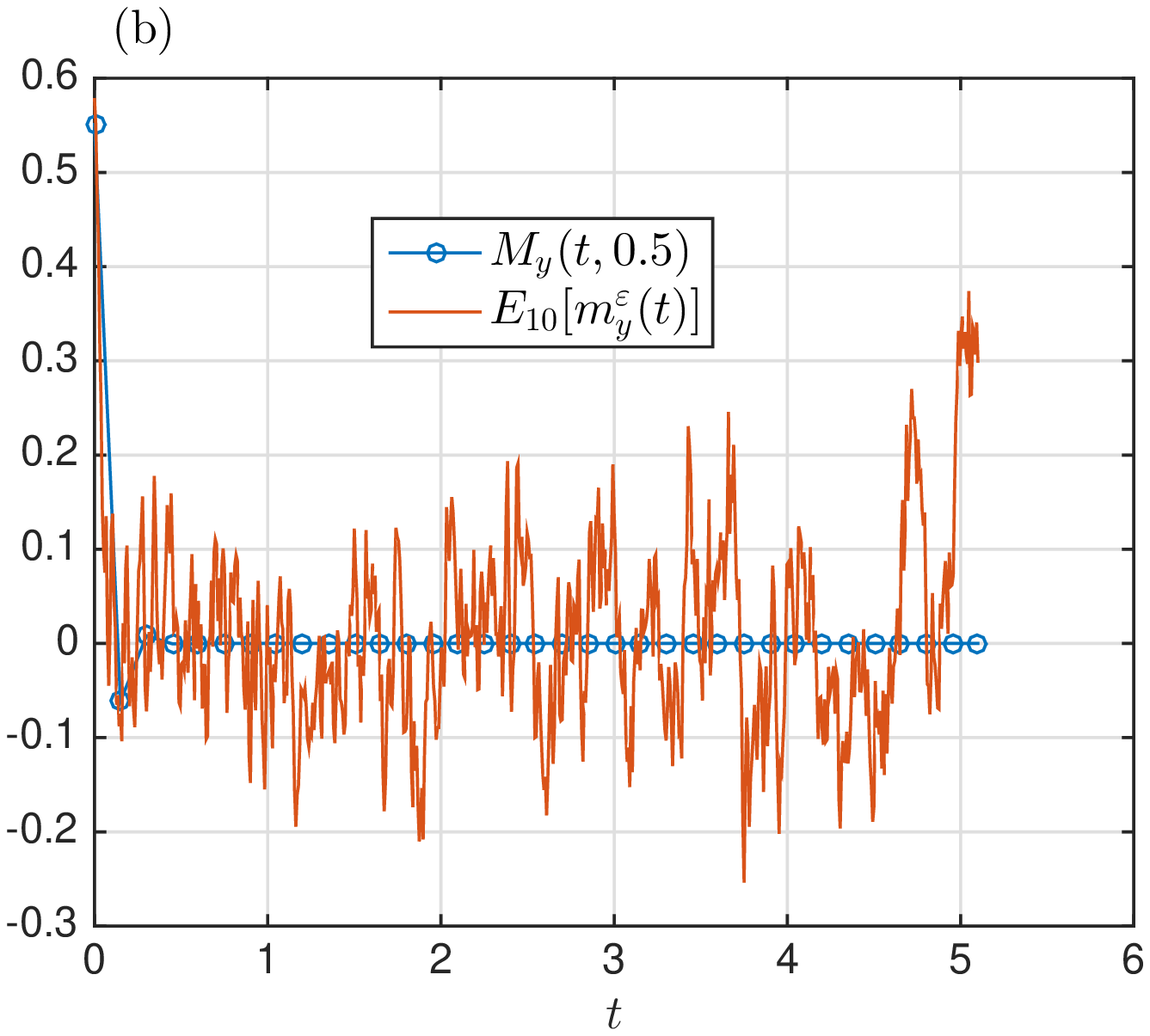}
    \includegraphics[width=0.47\textwidth]{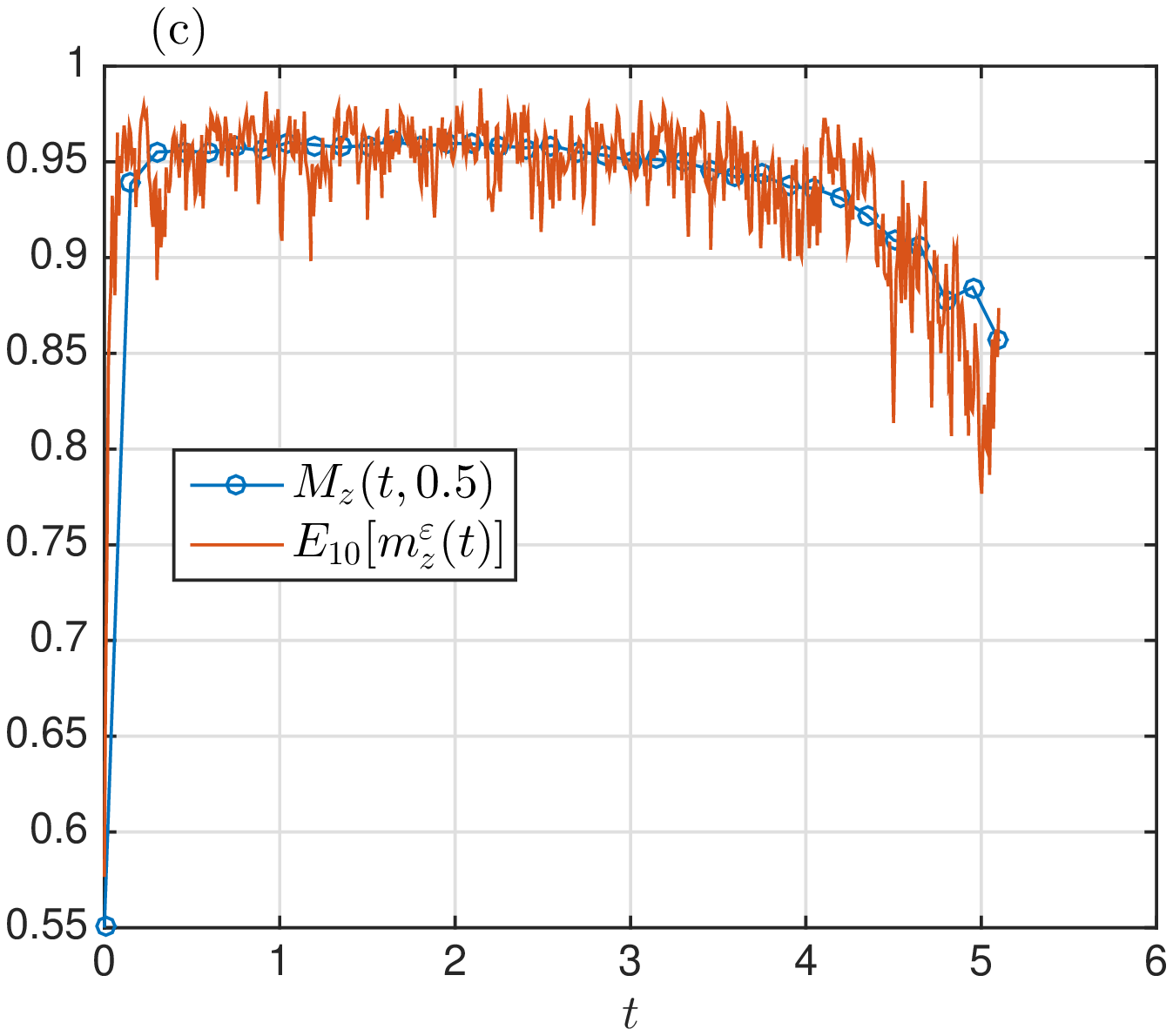}
    \includegraphics[width=0.47\textwidth]{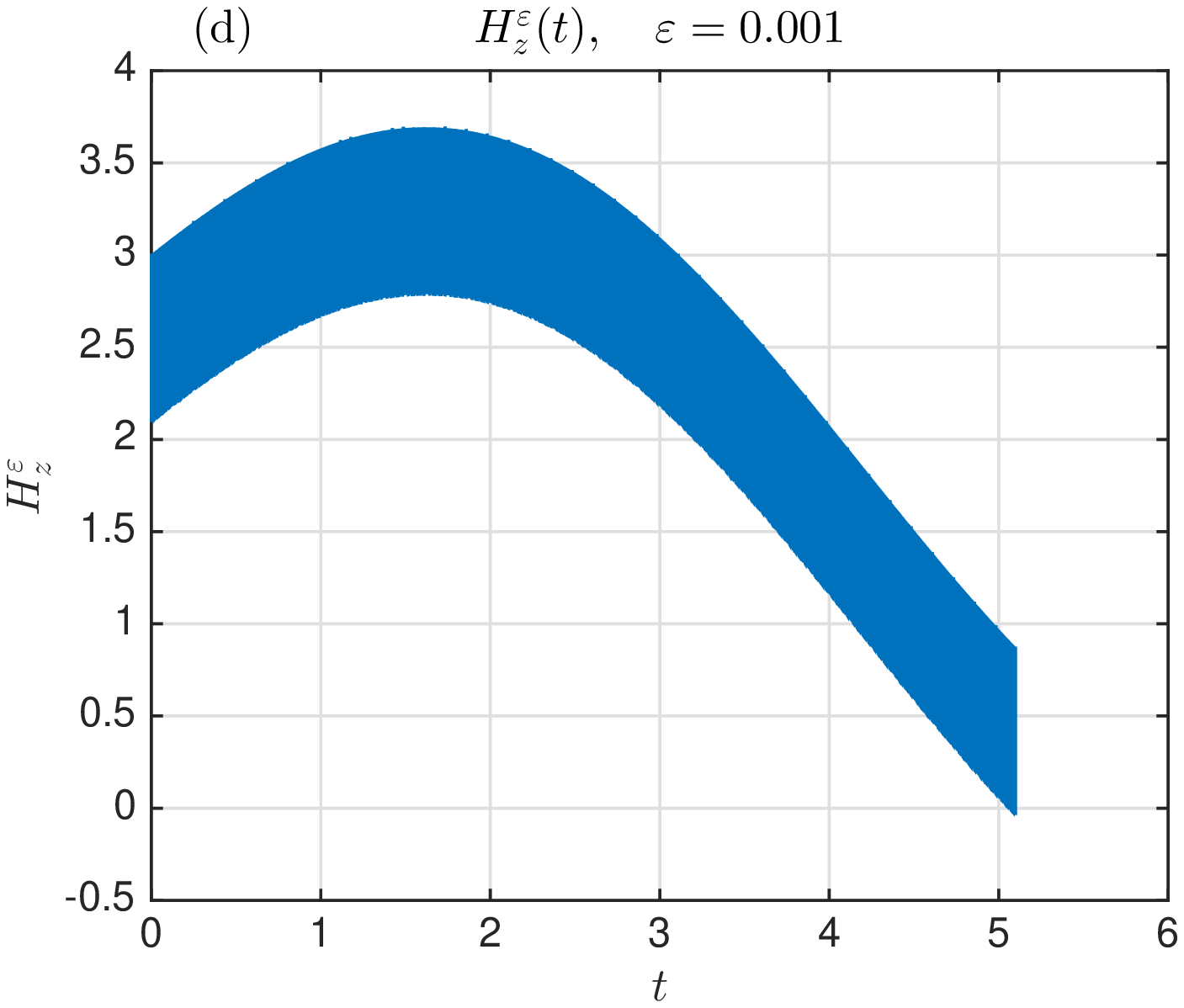}
  \end{center}
  \caption{1D chain temporal upscaling under an oscillatory in time and uniform in space external field. The $x$, $y$, $z$ components of the HMM solution and a direct numerical simulation computed by averaging the statistics of $10$ particles (a-c) in the presence of the external field $\vectorn{H}_{\mathrm{e}}^{\e}$ given by \eqref{eq:ExternalField_Uniform}. The value $\e = 0.001$ is used in these simulations. The $z$-component of the external field in \eqref{eq:ExternalField_Uniform} (d). }
  \label{fig:1DChain_Uniform}
\end{figure}

The second example concerns a case, where $10$ magnetic particles (defined on a one-dimensional equidistant grid $\{x_{i} \}_{i=0}^{9}$) are interacting. The parameter values $K_\mathrm{a} = 0$, $J_{ij}  = 1$ for $|i-j| \leq 1$, and $J_{ij} = 0$ when $|i-j| > 1$, and $\mu =  1$ are used in the simulation, and  periodic boundary conditions are used for the chain of particles. The number of macroscopic variables $\{ \vectorn{M}_{I} \}_{I=0}^{9}$ is equal to that of the microscopic variables $\{ \vectorn{m}^{\e}_{i} \}_{i=0}^{9}$, and the multiscaling is performed only in time. Namely, the micro problem \eqref{eq_Micromodel}, and the stochastic LLG equation \eqref{eq:LengthScaling} are solved globally (with the periodic boundary conditions), and the upscaling step \eqref{eq:Upscaling} uses averaging only in time. At first, the algorithm was applied to a case with an external field, which is aligned along the $z$-direction: 
\begin{equation}
  \label{eq:ExternalField_Uniform}
  \vectorn{H}_{\mathrm{e},i}^{\e}(t) = (1+\cos(0.43  t)+ \sin(0.73 t) + \cos(2\pi t/\e)^2)\vectorn{e}_z, \quad i=0,\ldots,9.
\end{equation}
This field is uniform in space and oscillatory in time. Similar to the example of a single spin, the HMM solution was compared to a direct numerical simulation. As the external field \eqref{eq:ExternalField_Uniform} is uniform in space, all the particles have the same statistical behaviour. Figure \ref{fig:1DChain_Uniform} shows that the magnetisation length is accurately captured for all the components. A decrease in the magnitude of the external field leads to a decrease in the magnetisation length as the behaviour of the system becomes more dominated by the thermal noise (leading to a disordered system). Note the decrease in the magnetisation length in Figure \ref{fig:1DChain_Uniform}c as the external field, shown in Figure \ref{fig:1DChain_Uniform}d, is decreasing. The notation $M_\nu(t,q)$ in the figure is used for the HMM solution at the spatial point $x=q$, where $\nu$ can be $x,y,$ or $z$ coordinate. For these simulations, an implicit mid-point rule with $35$ macroscopic time steps is used at the macroscopic scale and the final simulation time is $t = 5.1$. The Heun method \cite{Scholz2001} is used for timestepping in the micro problem. An interval of size $\tau = 5 \e$ is used for the micro problem. The physical parameters are chosen to be $\beta_\mathrm{L}= 1$, $\alpha_\mathrm{L} = 10$, and $D = 0.2$. 

\begin{figure}
  \begin{center}
    \includegraphics[width=0.47\textwidth]{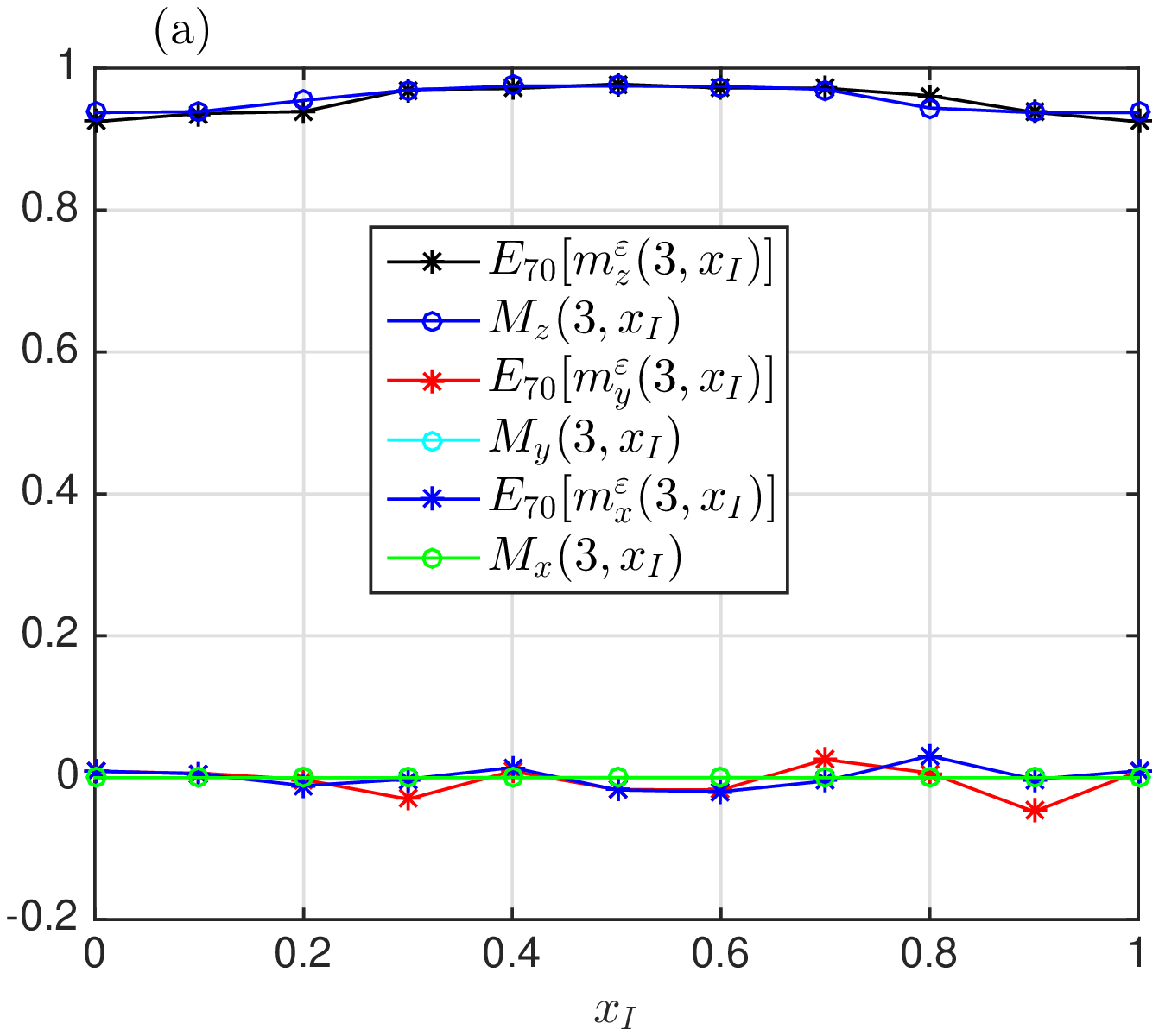}
    \includegraphics[width=0.47\textwidth]{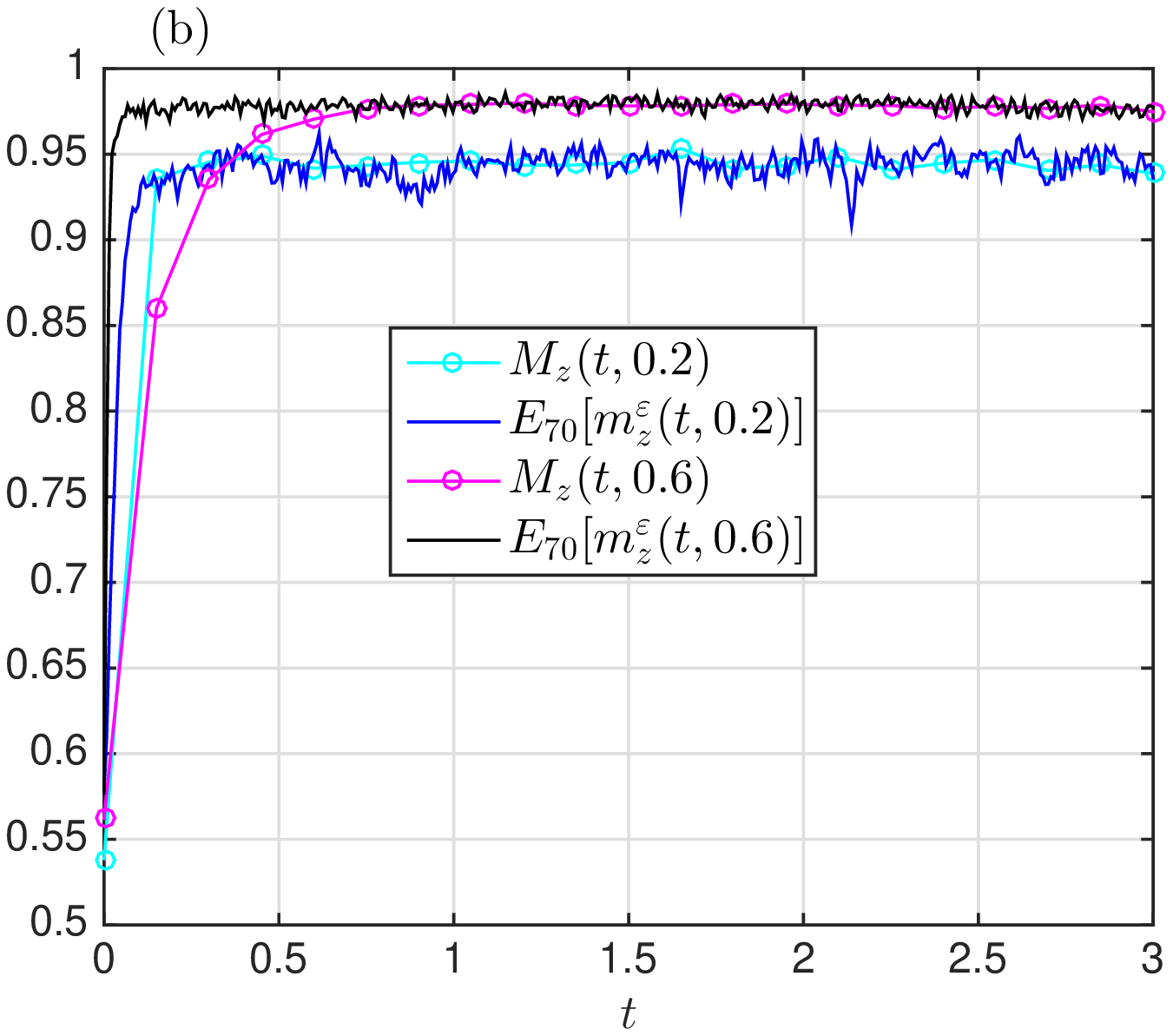}
    \includegraphics[width=0.47\textwidth]{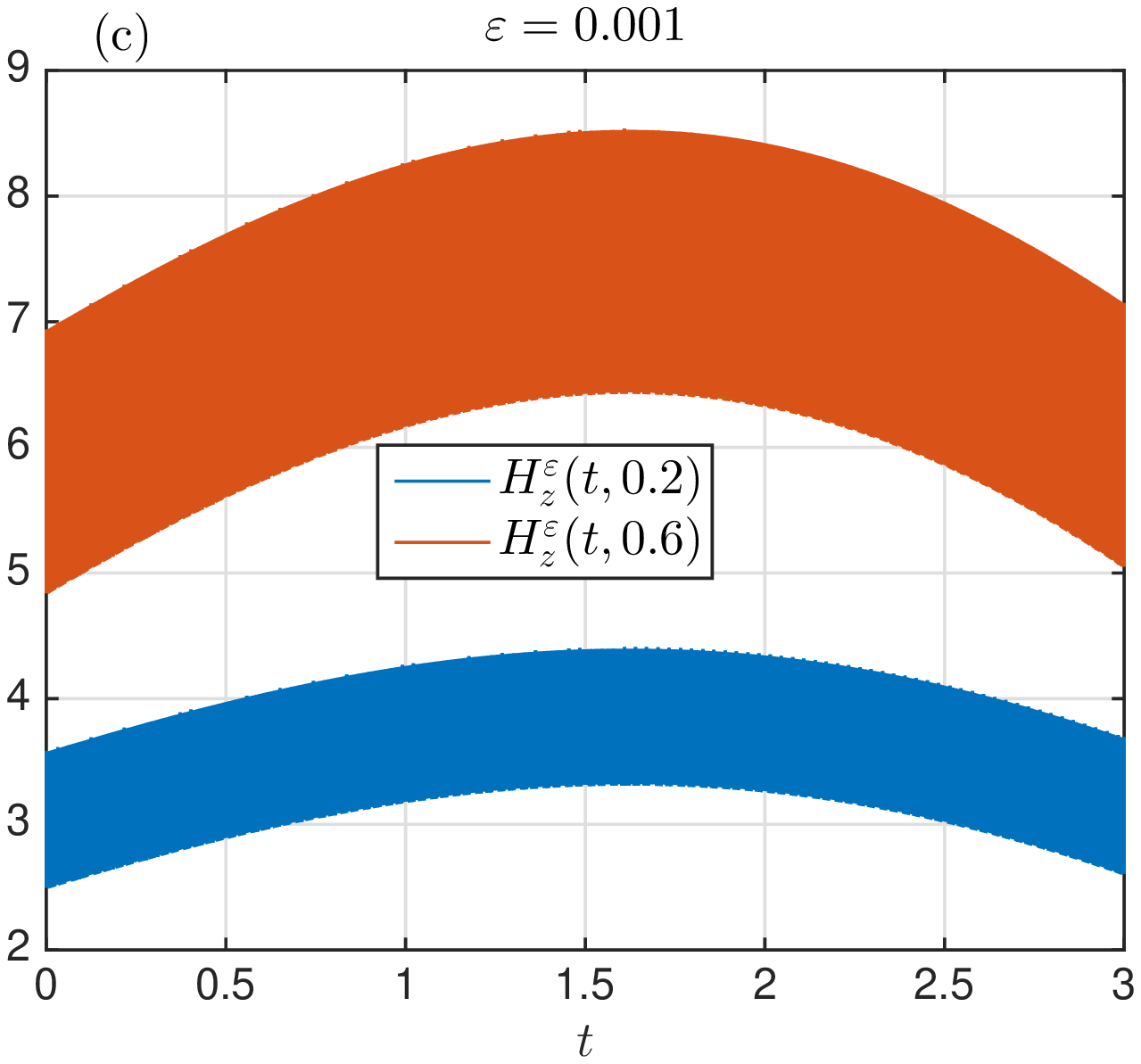}
  \end{center}
  \caption{1D chain temporal upscaling under an oscillatory in time and nonuniform in space external field. The HMM solutions in $x$,$y$,$z$-directions (a) are compared to a direct numerical simulation  computed by averaging 70 replicas of the stochastic LLG equation \eqref{eq:ASD_LLG} with the external field $\vectorn{H}_{\mathrm e}^{\e}$ given by \eqref{eq:ExternalField_NonUniform} (c). The time evolution of the magnetisation (in $z$-direction) at two different points in space (b).}
  \label{fig:1DChain_Nonuniform}
\end{figure}

Another example deals with the case when the external field is oscillatory in time but nonuniform in space. The molecular field approximation is not valid in this case. In particular, the external field is set to
\begin{equation}
  \label{eq:ExternalField_NonUniform}
  \vectorn{H}_{\mathrm{e},i}^{\e}(t) = (1+\cos(0.43  t)+ \sin(0.73 t) + \cos(2\pi t/\e)^2)(1.5 - \cos(2 \pi x_{i})) \vectorn{e}_z, \quad i=0,\ldots,9.
\end{equation}
When the external field is nonuniform in space, the magnetisation length should be nonuniform as well. Figure \ref{fig:1DChain_Nonuniform}b clearly shows that the magnetisation converges to different equilibrium states for two different points in space ($x = 0.2$ and $x=0.6$). The corresponding external fields at these two points are shown in \ref{fig:1DChain_Nonuniform}c. In these simulations $20$ macroscopic time steps are used, and the final simulation time is $t=3$. All other parameter values are the same as in the uniform field case shown in Figure \ref{fig:1DChain_Uniform}. The HMM is shown to capture the correct magnetisation dynamics for this example of a non-uniform external field. Note that the expected value of the exact reference solution (direct numerical solution) is computed by averaging 70 replicas of the solution. In Figure \ref{fig:1DChain_Nonuniform}a, a small deviation between the HMM solution and the reference solution is observed. This is mainly due to the fact that only $70$ replicas are used while approximating the exact expected value. These deviations decrease upon computing the expected value using a larger number of copies. 

\subsubsection{HMM: Chain of spins, multiscaling in time and space}
\label{SubSub_TimeSpaceHMM}

\begin{figure}
  \begin{center}
    \includegraphics[width=0.47\textwidth]{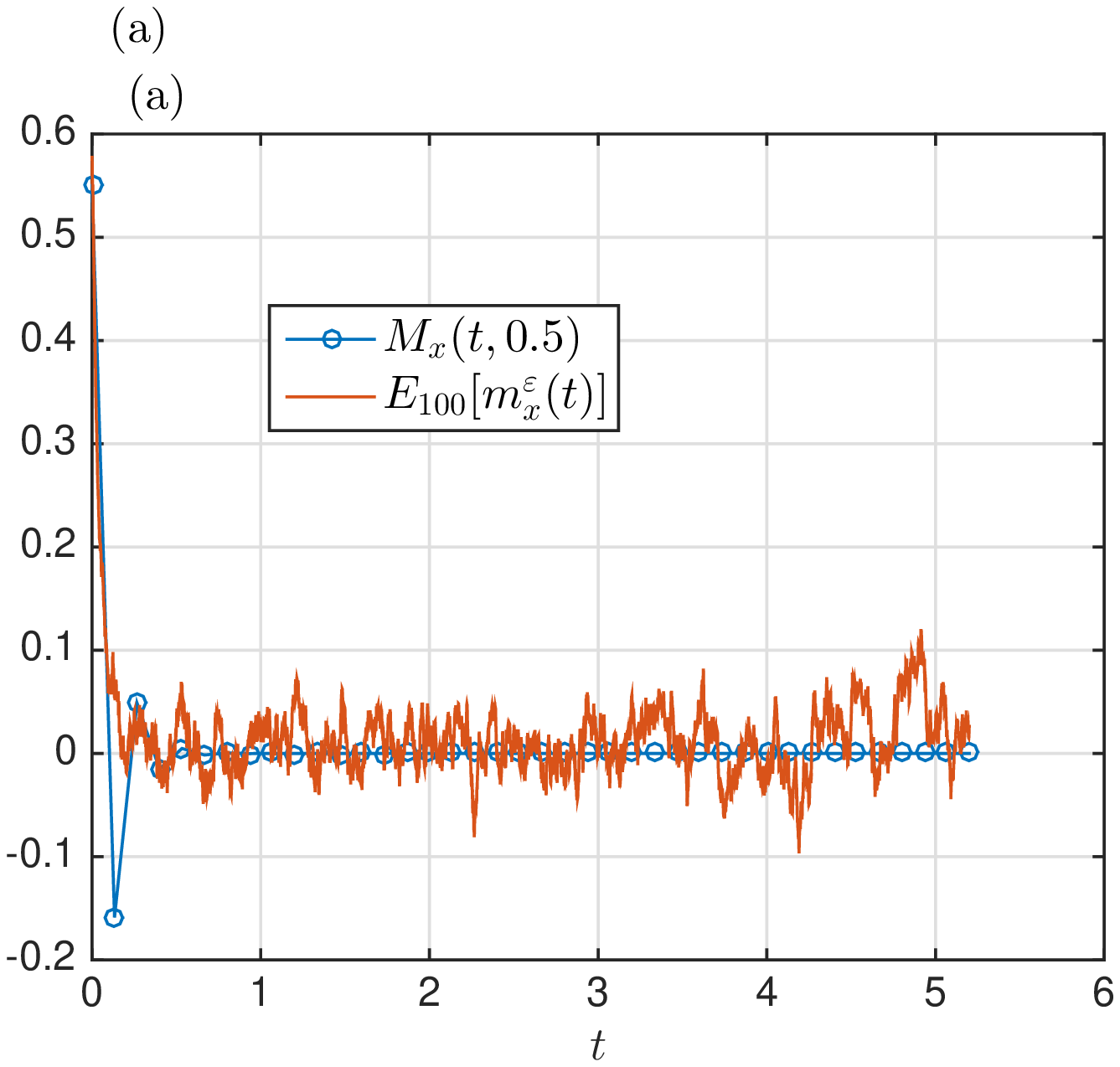}
    \includegraphics[width=0.47\textwidth]{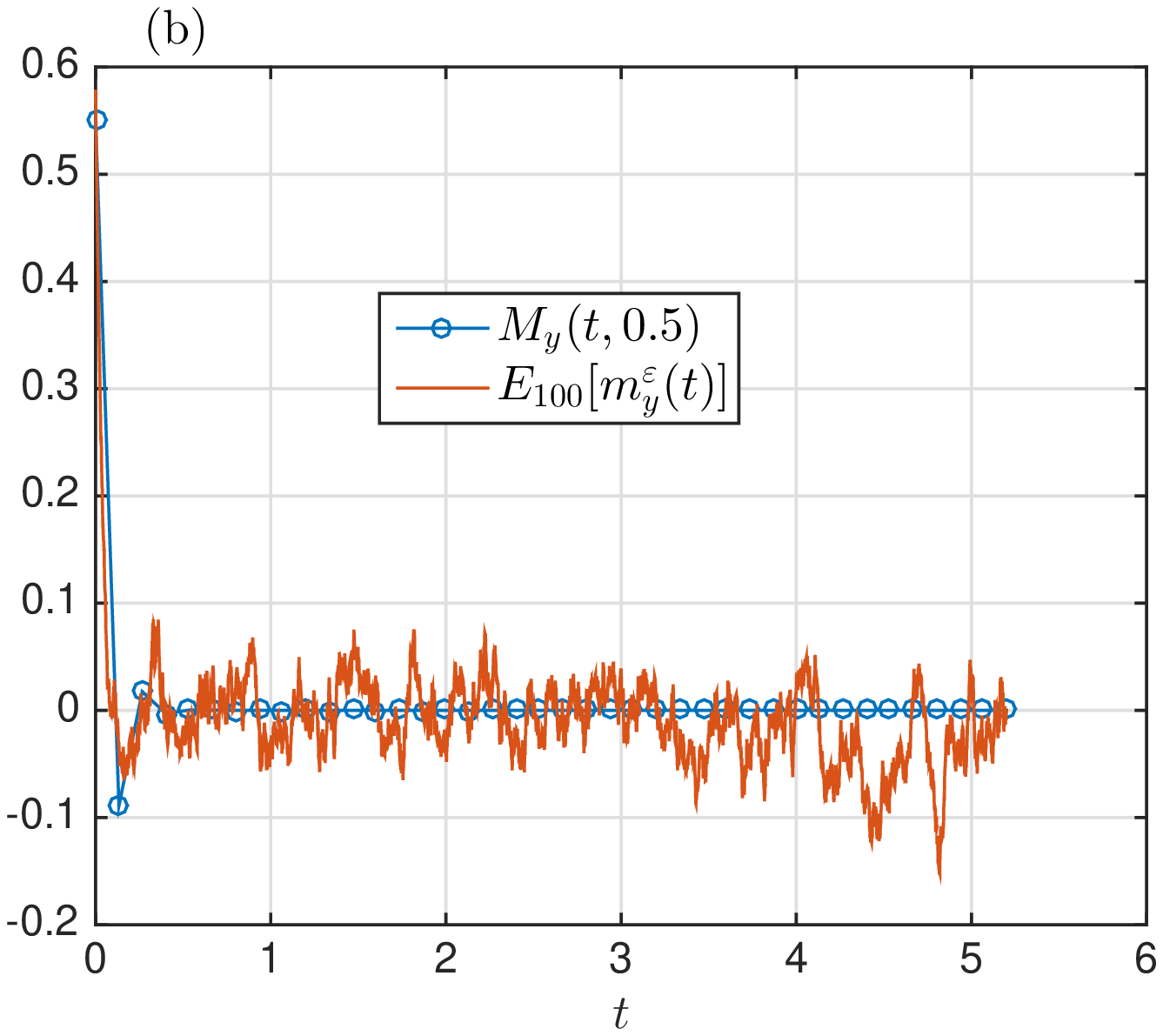}
    \includegraphics[width=0.47\textwidth]{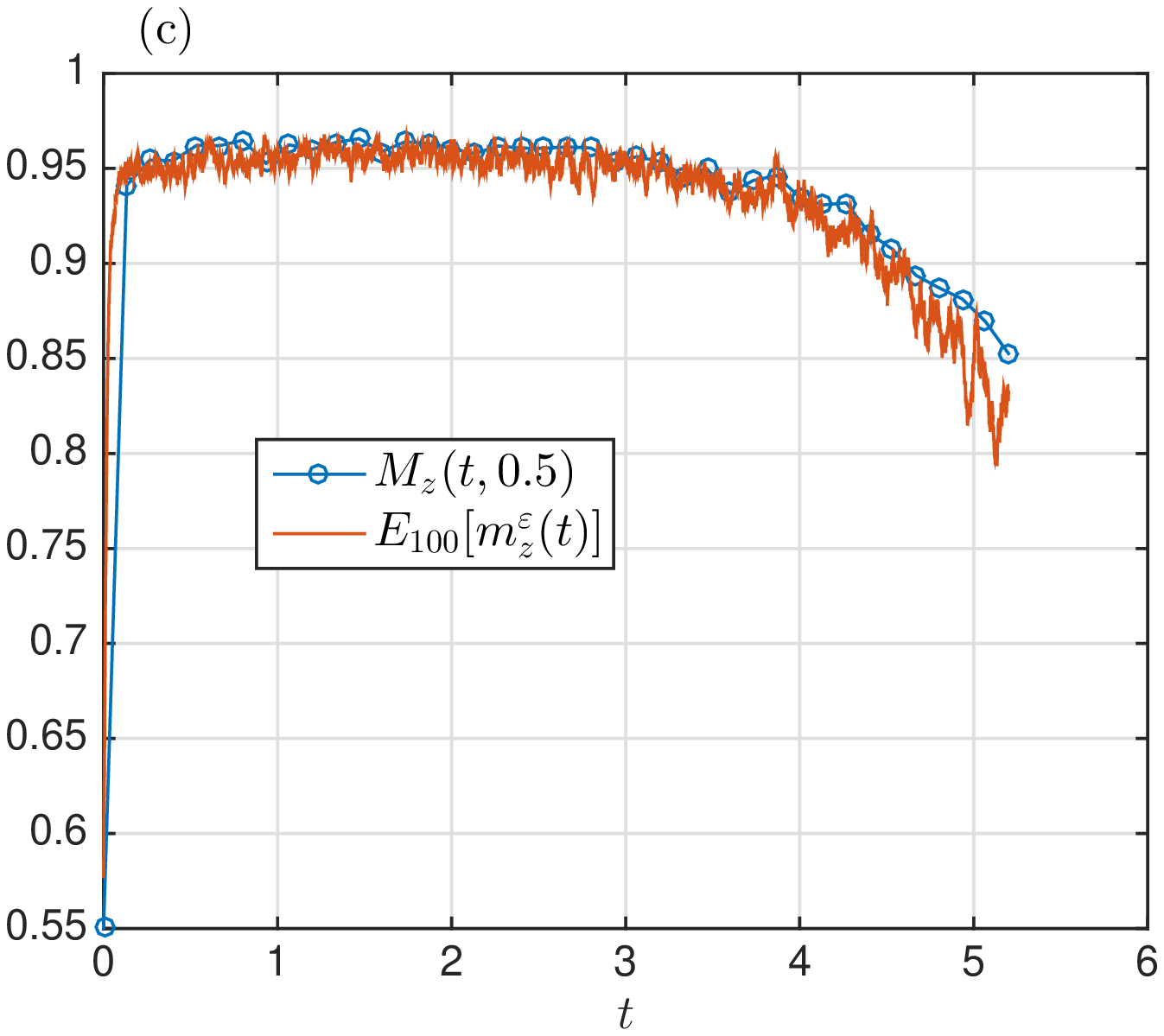}
  \end{center}
  \caption{1D chain temporal upscaling under an oscillatory in time and uniform in space external field. The HMM solutions in $x$,$y$,$z$-directions are compared to the exact magnetisation computed by averaging the statistics of 100 magnetic moments (a-c) with the external field $\vectorn{H}_{\mathrm e}^{\e}$ given by \eqref{eq:ExternalField_Uniform}. The following parameters are chosen for the simulation: $\eta = 0.1$, $\e = 0.001, \tau = 5 \e$, $D = 0.2$. }
  \label{fig:1DChain_Nonuniform_FullHMM}
\end{figure}

In this example, the case of a chain of magnetic particles, which are subjected to the external field \eqref{eq:ExternalField_Uniform}, is considered. The atomistic particles are defined on a fine grid $\{ x_{i} \}_{i=0}^{99}$, whereas the macroscopic magnetisations are defined on a coarse grid $\{ X_{I} \}_{I=0}^{9}$. Moreover, each macroscopic variable corresponds to an average of $11$ magnetic moments. Here, the HMM uses upscaling both in time and space.  Figure \ref{fig:1DChain_Nonuniform_FullHMM} confirms that the full HMM algorithm (namely multiscaling in time and space) performs equally well. The number of time steps that were used in the macro model was $35$, while other numerical parameter values used for these simulations are $\eta = 0.1$ (or $r = 5$ which means that the macroscopic variables are computed by averaging $2r+1 = 11$ spins), $\e = 0.001, \tau = 5 \e$, and the physical parameters are $\beta_\mathrm{L}=1$, $\alpha_\mathrm{L}=10$, $D = 0.2$, $J_{ij}  = 1$ for $|i-j| \leq 1$, and $J_{ij} = 0$ when $|i-j| > 1$, and $\mu = 1$. The macroscopic solver uses an implicit mid-point rule with $40$ time steps and the dynamics up until $t=5.2$ is shown. Similar to previous examples, the Heun method is used as the microscopic solver. 

\subsubsection{Limitations of HMM}

\begin{figure}
  \begin{center}
    \includegraphics[width=0.47\textwidth]{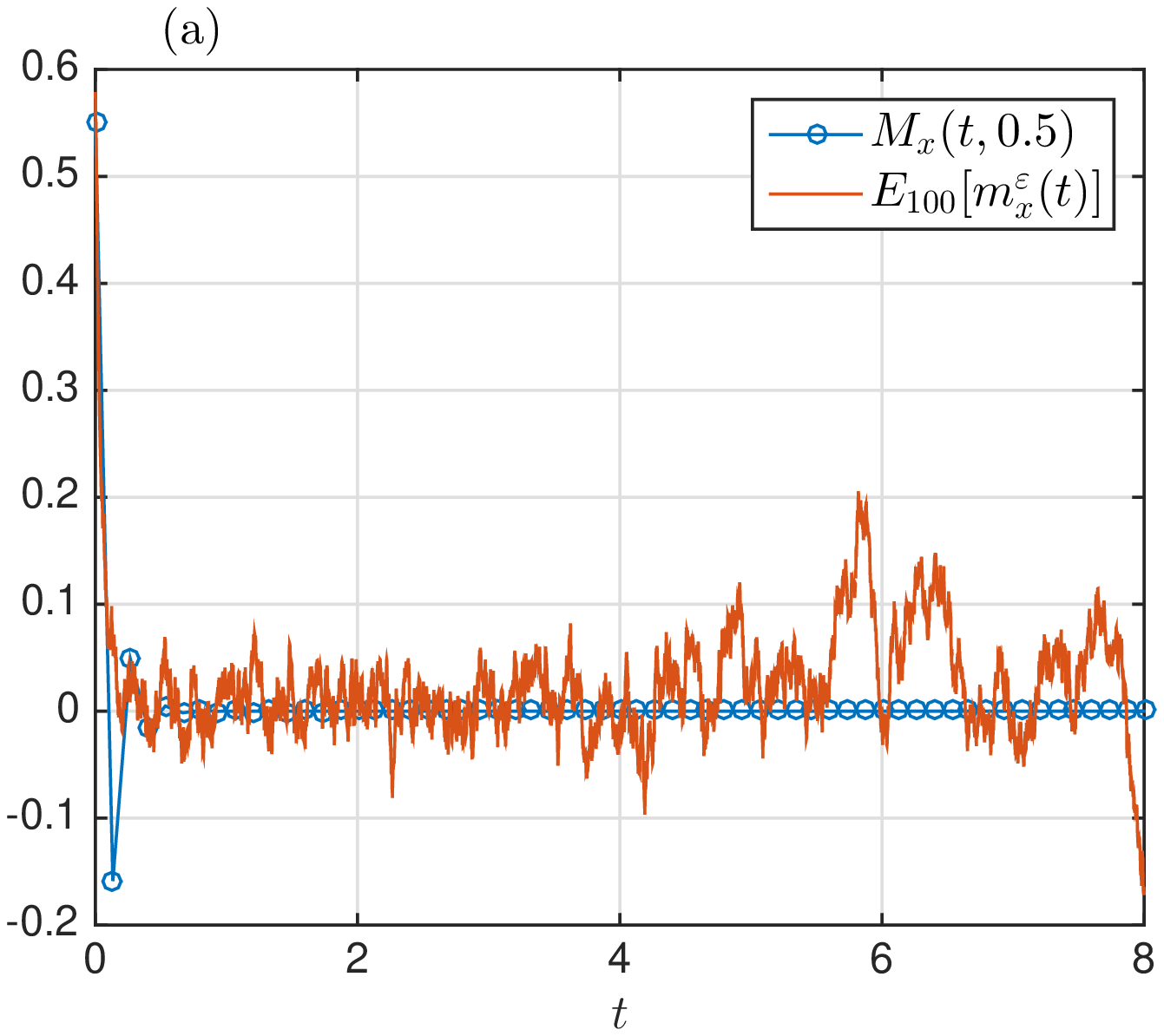}
    \includegraphics[width=0.47\textwidth]{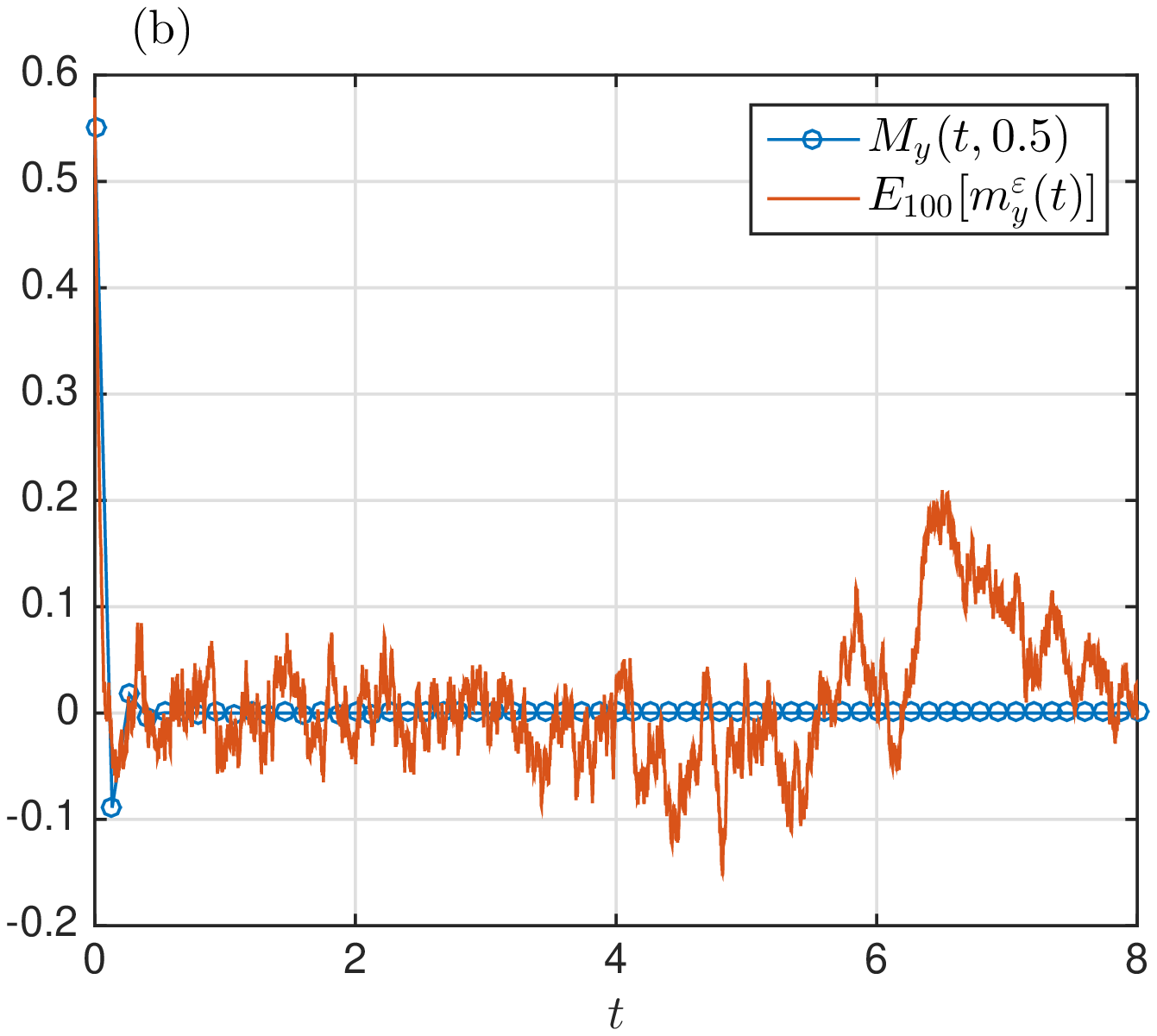}
    \includegraphics[width=0.47\textwidth]{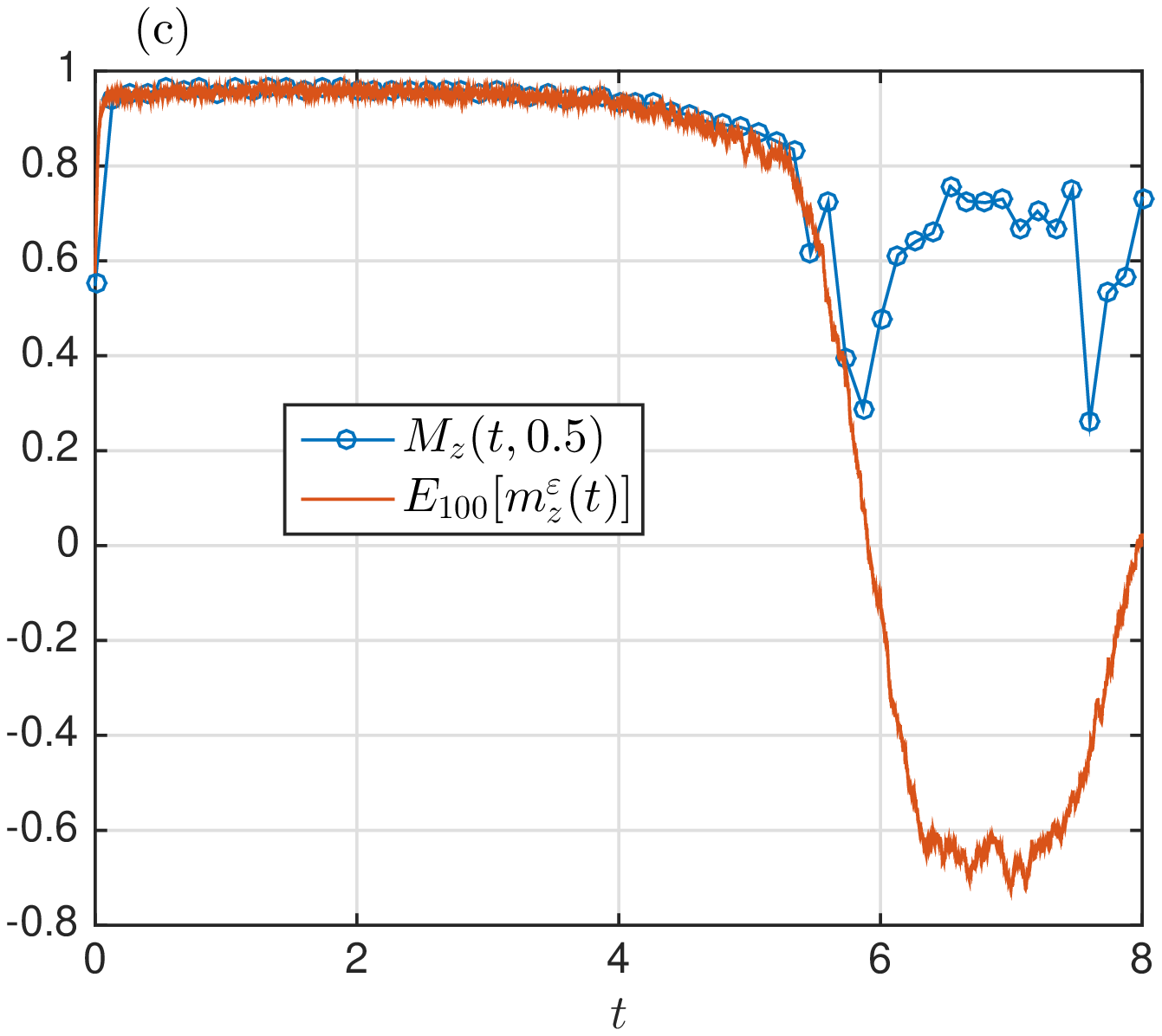}
    \includegraphics[width=0.47\textwidth]{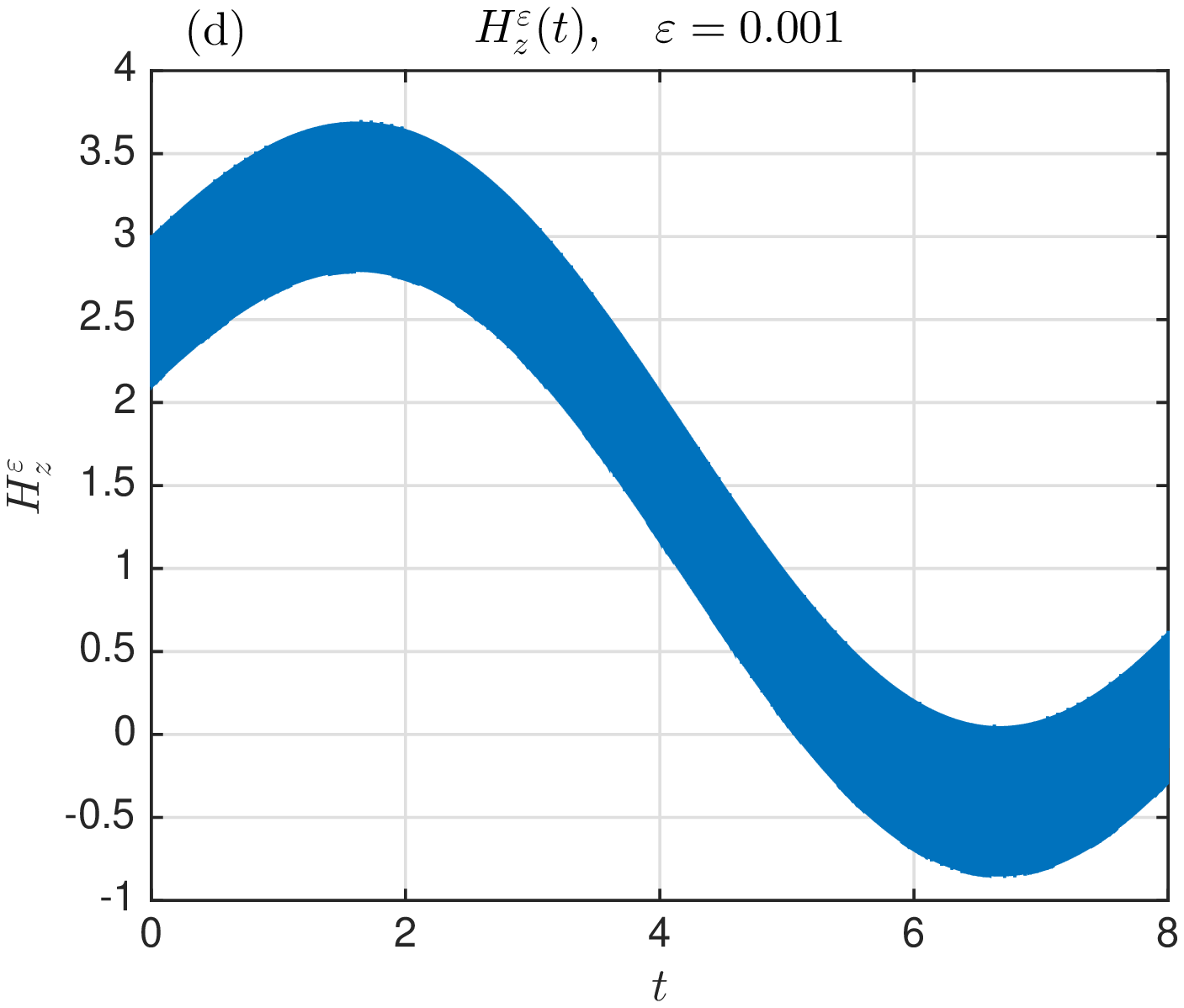}
  \end{center}
  \caption{1D chain temporal upscaling under an oscillatory in time and uniform in space external field. The HMM solutions in $x$,$y$,$z$-directions are compared to the exact magnetisation computed by averaging  the statistics of 100 magnetic moments (a-c)  with the external field $\vectorn{H}_{\mathrm e}^{\e}$ given by \eqref{eq:ExternalField_Uniform}. The following parameters are chosen for the simulation: $\eta = 0.1$, $\e = 0.001, \tau = 5 \e$, $D = 0.2$. }
  \label{fig:1DChain_Uniform_FullHMMLong}
\end{figure}

In subsection \ref{SubSub_TimeSpaceHMM}, the dynamics of the magnetisation was simulated until $t=5.2$. Here, precisely the same example with a longer simulation time, $t=8$, is considered. When $t \approx 6$ the external field \eqref{eq:ExternalField_Uniform} oscillates around zero, see Figure \ref{fig:1DChain_Uniform_FullHMMLong}d. In this regime, the system becomes dominated by the noise, as external field is too small to stabilise the system, and a large number of magnetic moments is required to capture the correct statistics, see Figure \ref{fig:Asize}. This fact leads to a breakdown of the HMM strategy around $t\approx 6$, see Figure \ref{fig:1DChain_Uniform_FullHMMLong}a-c. If the number of particles in the microscopic simulations is large, the cost of the HMM will not differ from that of a full atomistic simulation. Hence, switching to the full atomistic simulations is preferable, in a short temporal regime around $t \approx 6$, and once the system is stable enough, the HMM algorithm may be employed again.

\section{Conclusions}
\label{sec:conclusions}

In this article, application of atomistic-continuum multiscale computational strategies to modelling finite-temperature magnetisation dynamics was discussed. There is a number of issues that arise due to additional noise term appearing at the atomistic scale, which takes into account non-zero temperature condition. In particular, when the ensemble average behaviour of the atomistic model (ASD) is obtained, the result cannot be accurately described using simple continuum models, as they rely on different restrictive assumptions regarding magnetisation length. This inconsistency creates problems for coupling techniques.

Two major coupling techniques were discussed in this article: domain partitioning and upscaling via HMM. These techniques aim at modelling somewhat different cases, but the approaches are complementary to each other. Domain partitioning is suitable when the difference between spatial scales, which are relevant for modelling certain phenomena, is not large. Moreover, as it is explicitly required to use a \emph{specific continuum model} at the continuum region in the case of domain partitioning, the application of this approach is limited to relatively low temperatures, since at high temperatures, non-linear dependency of the continuum parameters on magnetisation gradient, which is not taken into account in established models, plays a significant role. The selection of a continuum model for the case of domain partitioning was further discussed in Section \ref{sec:model_sel}.

It was shown that by using a damping band it is possible to filter out some high-frequency noise at the atomistic-continuum interface. However, by design, this damping band cannot filter out low frequency waves, which originate in the atomistic region and some of which are the result of thermal fluctuations. This leads to inhomogeneous in space reduction of magnetisation length in the continuum region, when ensemble averages are calculated. Such behaviour is certainly unphysical and is the major problem for domain partitioning approach when the continuum region is dynamic. Depending on the magnitude of the temperature and timescales involved in the modelling, this artifact can be neglected, but it fully disappears only when the continuum model is quasistatic.

Upscaling-based approaches are well-suited for problems with large separation of scales. Such a strategy is proposed under the HMM framework. A microscale model, describing the dynamics of spin magnetic moments in spatially confined regions, is used to obtain parameters needed to close an initially incomplete macroscale model. The method uses a two-way coupling between a macroscopic and a microscopic model. The macroscopic solver uses large step sizes in time and space, leading to a significant reduction of the degrees of freedom to be computed in comparison to a full atomistic spin dynamics simulation. An analysis for the computational cost is given in Remark \ref{Rem:Cost_Alg1} for a single spin at zero-temperature. It is shown that the HMM requires a much lower computational cost in comparison to a direct numerical simulation. Moreover, similar arguments can be safely made when interactions between the atomistic particles are present. The main computational gain comes from the fact that the atomistic model is solved only locally both in time and space, while a direct numerical simulation entails a full resolution of the microscopic scales over the entire temporal and spatial domain. The method has  been observed to capture the correct magnetisation dynamics in three different examples. 

The upscaling-based approach, which is presented here, is complementary to the domain partitioning approach in the sense that the continuum model, which is required for the domain partitioning, can be obtained by the upscaling, rather than using other physically restrictive continuum models. While using such upscaling methods, the limitations should also be taken into account. For systems which are  dominated by noise (weak external field and/or anisotropy), a large number of atomistic spins is required for an accurate representation of macroscopic scale quantities. This is a case, where HMM would require a computational cost similar to that of a full atomistic simulation.

\section*{Acknowledgements}

The authors would like to acknowledge the support of eSSENCE. The authors would also like to acknowledge the support from the Swedish Research Council (VR) and the KAW foundation (grants 2013.0020 and 2012.0031).

\section{Appendix}
\label{Sec:Appendix}

Although the mean-field approximation is well-known and is presented elsewhere \cite{Aharoni1996}, different notation is used throughout literature. Therefore, the key concepts are repeated here.

\subsection{A single spin interacting with a magnetic field}

When a single spin $\vectorn{m}$ interacts with an external field $\vectorn{H} = H \vectorn{e}_z$, the total energy can be written as 
\begin{equation*}
  E = - \vectorn{m} \cdot \vectorn{H} , \quad \left| \vectorn{m} \right| = S ,
\end{equation*}
and the spin alignes with the field; however, the statistical average of $z$-componet of $\vectorn{m}$ is given by \cite{Aharoni1996}:
\begin{equation*}
  \left\langle m_z \right\rangle = \frac{\int_{-S}^{S} s e^{s \gamma} \; \mathrm{d}s}{\int_{-S}^{S} e^{s \gamma} \; \mathrm{d}s} , \quad
  \gamma = \frac{H}{k_{\mathrm{B}} T} . 
\end{equation*}
After integrals are calculated, $m_z$ is given by
\begin{equation*}
  \left\langle m_z \right\rangle = S \coth\left( S \gamma \right) - \frac{1}{\gamma} = S L \left( S \gamma \right) ,
\end{equation*}
where $L(x)$ is the Langevin function.

\subsection{A chain of interacting magnetic particles  subjected to an external magnetic field}

When a number of interacting magnetic particles $\vectorn{m}_i$ are subjected to a uniform external field $\vectorn{H}_\mathrm{e}$, the total energy\footnote{Here the total energy formulation follows \cite{Evans2014}; however other formulations exist.} is given by 
\begin{equation*}
  E = - \frac{1}{2} \sum_{i} \sum_{\substack{j\\j \neq i}} J_{ij} \vectorn{m}_i \cdot \vectorn{m}_j  - \frac{1}{2} \sum_i K_\mathrm{a} \left( \vectorn{p}_\mathrm{a} \cdot \vectorn{m}_i \right)^2 - \sum_i \mu \vectorn{m}_i \cdot \vectorn{H}_\mathrm{e} , \quad \left| \vectorn{m}_k \right| = 1, \quad \forall k .
\end{equation*}
In a mean-field approximation, the contributions of individual spins $\vectorn{m}_i$ to the total energy are separated and for each $\vectorn{m}_i$ all neighbouring moments are replaced by their statistical averages in the following way:
\begin{equation*}
  E_{i} = - \sum_{\substack{j\\j \neq i}} J_{ij} \vectorn{m}_i \cdot \left\langle \vectorn{m}_j \right\rangle - K_\mathrm{a} \vectorn{m}_i \cdot \vectorn{p}_\mathrm{a} \vectorn{p}_\mathrm{a} \cdot \left\langle \vectorn{m}_i \right\rangle - \mu \vectorn{m}_{i} \cdot \vectorn{H}_\mathrm{e} = - \vectorn{m}_{i} \cdot \vectorn{H}_{i} .
\end{equation*}
Moreover, a simple case of $\vectorn{H}_\mathrm{e} = H_\mathrm{e} \vectorn{e}_z$ and $\vectorn{p}_\mathrm{a} = \vectorn{e}_z$ is usually considered, which means that all statistical averages of all spins should be aligned along $\vectorn{e}_z$. In this case, it is additionally assumed that the statistical averages of all the spins are equal, i.e. $\left\langle m_i^{z} \right\rangle = \left\langle m_j^{z} \right\rangle$. Using the derivation from the previous section, $\left\langle m_i^{z} \right\rangle$ is obtained:
\begin{equation*}
  \left\langle m_i^{z} \right\rangle = L \left( \frac{1}{k_{\mathrm{B}} T} \left( \left\langle m_i^{z} \right\rangle \sum_{\substack{j\\j \neq i}} J_{ij} + \left\langle m_i^{z} \right\rangle K_\mathrm{a} + \mu H_\mathrm{e} \right) \right) .
\end{equation*}

\bibliographystyle{spmpsci}
\bibliography{refs}

\begin{thebibliography}{10}
\providecommand{\url}[1]{{#1}}
\providecommand{\urlprefix}{URL }
\expandafter\ifx\csname urlstyle\endcsname\relax
  \providecommand{\doi}[1]{DOI~\discretionary{}{}{}#1}\else
  \providecommand{\doi}{DOI~\discretionary{}{}{}\begingroup
  \urlstyle{rm}\Url}\fi

\bibitem{Abdulle_E_Engquist2012}
Abdulle, A., E, W., Engquist, B., Vanden-Eijnden, E.: {T}he heterogeneous
  multiscale method.
\newblock Acta Numerica \textbf{21}, 1--87 (2012).
\newblock \doi{10.1017/S0962492912000025}

\bibitem{Aharoni1996}
Aharoni, A.: {I}ntroduction to the theory of ferromagnetism.
\newblock Oxford University Press (1996)

\bibitem{Andreas2014}
Andreas, C., K\'{a}kay, A., Hertel, R.: {M}ultiscale and multimodel simulation
  of {B}loch-point dynamics.
\newblock Physical Review B \textbf{89}(13), 134,403 (2014).
\newblock \doi{10.1103/PhysRevB.89.134403}

\bibitem{Arjmand2016}
Arjmand, D., Engblom, S., Kreiss, G.: {T}emporal upscaling in micromagnetism
  via heterogeneous multiscale methods  (2016).
\newblock ArXiv:1603.04920

\bibitem{Arjmand_Runborg2014}
Arjmand, D., Runborg, O.: {A}nalysis of heterogeneous multiscale methods for
  long time wave propagation problems.
\newblock Multiscale Modeling \& Simulation \textbf{12}(3), 1135--1166 (2014).
\newblock \doi{10.1137/140957573}

\bibitem{Arjmand_Runborg2016a}
Arjmand, D., Runborg, O.: {A} time dependent approach for removing the cell
  boundary error in elliptic homogenization problems.
\newblock Journal of Computational Physics \textbf{314}, 206--227 (2016).
\newblock \doi{10.1016/j.jcp.2016.03.009}

\bibitem{Atxitia2010}
Atxitia, U., Hinzke, D., Chubykalo-Fesenko, O., Nowak, U., Kachkachi, H.,
  Mryasov, O.N., Evans, R.F., Chantrell, R.W.: {M}ultiscale modeling of
  magnetic materials: {T}emperature dependence of the exchange stiffness.
\newblock Physical Review B \textbf{82}(13), 134,440 (2010).
\newblock \doi{10.1103/PhysRevB.82.134440}

\bibitem{Banas_Prohl2014}
Banas, L., Brzezniak, Z., Neklyudov, M., Prohl, A.: {S}tochastic
  ferromagnetism: {A}nalysis and numerics.
\newblock De Gruyter (2013)

\bibitem{Bergqvist2013}
Bergqvist, L., Taroni, A., Bergman, A., Etz, C., Eriksson, O.: {A}tomistic spin
  dynamics of low-dimensional magnets.
\newblock Physical Review B \textbf{87}(14), 144,401 (2013).
\newblock \doi{10.1103/PhysRevB.87.144401}

\bibitem{Brown1963}
Brown, W.F.: {M}icromagnetics.
\newblock Interscience Publishers (1963)

\bibitem{Brown1963b}
Brown, W.F.: {T}hermal fluctuations of a single-domain particle.
\newblock Physical Review \textbf{130}(5), 1677--1686 (1963).
\newblock \doi{10.1103/PhysRev.130.1677}

\bibitem{Cervera2007}
Cervera, C.J.G.: {N}umerical micromagnetics: A review.
\newblock Bol. Soc. Esp. Mat. Apl. \textbf{39}, 103--135 (2007)

\bibitem{ChubykaloFesenko2006}
Chubykalo-Fesenko, O., Nowak, U., Chantrell, R.W., Garanin, D.: {D}ynamic
  approach for micromagnetics close to the {C}urie temperature.
\newblock Physical Review B \textbf{74}(9), 094,436 (2006).
\newblock \doi{10.1103/PhysRevB.74.094436}

\bibitem{Cimrak2008}
Cimr\'{a}k, I.: {A} survey on the numerics and computations for the
  {L}andau-{L}ifshitz equation of micromagnetism.
\newblock Archives of Computational Methods in Engineering \textbf{15}(3),
  277--309 (2008).
\newblock \doi{10.1007/s11831-008-9021-2}

\bibitem{dAquino2005}
d'Aquino, M., Serpico, C., Miano, G.: {G}eometrical integration of
  {L}andau-{L}ifshitz-{G}ilbert equation based on the mid-point rule.
\newblock Journal of Computational Physics \textbf{209}(2), 730--753 (2005).
\newblock \doi{10.1016/j.jcp.2005.04.001}

\bibitem{E_Engquist2003}
E, W., Engquist, B.: {T}he heterogeneous multiscale methods.
\newblock Communications in Mathematical Sciences \textbf{1}(1), 87--132 (2003)

\bibitem{Eriksson2016}
Eriksson, O., Bergman, A., Bergqvist, L., Hellsvik, J.: {A}tomistic spin
  dynamics: {F}oundations and applications.
\newblock Oxford University Press (2016)

\bibitem{Evans2014}
Evans, R.F.L., Fan, W.J., Chureemart, P., Ostler, T.A., Ellis, M.O.A.,
  Chantrell, R.W.: {A}tomistic spin model simulations of magnetic
  nanomaterials.
\newblock Journal of Physics: Condensed Matter \textbf{26}(10), 103,202 (2014).
\newblock \doi{10.1088/0953-8984/26/10/103202}

\bibitem{Garanin1997}
Garanin, D.A.: {F}okker-{P}lanck and {L}andau-{L}ifshitz-{B}loch equations for
  classical ferromagnets.
\newblock Physical Review B \textbf{55}(5), 3050--3057 (1997).
\newblock \doi{10.1103/PhysRevB.55.3050}

\bibitem{GarciaSanchez2005}
Garcia-Sanchez, F., Chubykalo-Fesenko, O., Mryasov, O., Chantrell, R.W.,
  Guslienko, K.Y.: {E}xchange spring structures and coercivity reduction in
  {F}e{P}t/{F}e{R}h bilayers: {A} comparison of multiscale and micromagnetic
  calculations.
\newblock Applied Physics Letters \textbf{87}(12), 122,501 (2005).
\newblock \doi{10.1063/1.2051789}

\bibitem{Hinzke2008}
Hinzke, D., Kazantseva, N., Nowak, U., Mryasov, O.N., Asselin, P., Chantrell,
  R.W.: {D}omain wall properties of {F}e{P}t: {F}rom {B}loch to linear walls.
\newblock Physical Review B \textbf{77}(9), 094,407 (2008).
\newblock \doi{10.1103/PhysRevB.77.094407}

\bibitem{Jourdan2008}
Jourdan, T., Marty, A., Lan\c{c}on, F.: {M}ultiscale method for {H}eisenberg
  spin simulations.
\newblock Physical Review B \textbf{77}(22), 224,428 (2008).
\newblock \doi{10.1103/PhysRevB.77.224428}

\bibitem{Kevrikidis_2003}
Kevrekidis, I.G., Gear, C.W., Hyman, J.M., Kevrekidis, P.G., Runborg, O.,
  Theodoropoulos, C.: {E}quation-free, coarse-grained multiscale computation:
  {E}nabling microscopic simulators to perform system-level analysis.
\newblock Communications in Mathematical Sciences \textbf{1}(4), 715--762
  (2003)

\bibitem{Kirschner2005}
Kirschner, M.: {C}oarse-graining in micromagnetics.
\newblock Ph.D. thesis, Vienna University of Technology (2005)

\bibitem{Landau2005}
Landau, D.P., Binder, K.: {A} guide to {M}onte {C}arlo simulations in
  statistical physics.
\newblock Cambridge University Press (2009)

\bibitem{Miller2009}
Miller, R.E., Tadmor, E.B.: {A} unified framework and performance benchmark of
  fourteen multiscale atomistic/continuum coupling methods.
\newblock Modelling and Simulation in Materials Science and Engineering
  \textbf{17}(5), 053,001 (2009).
\newblock \doi{10.1088/0965-0393/17/5/053001}

\bibitem{Poluektov2016b}
Poluektov, M., Eriksson, O., Kreiss, G.: {C}oupling atomistic and continuum
  modelling of magnetism  (2016).
\newblock Submitted

\bibitem{Poluektov2016}
Poluektov, M., Eriksson, O., Kreiss, G.: {S}cale transitions in magnetisation
  dynamics.
\newblock Communications in Computational Physics \textbf{20}(4), 969--988
  (2016).
\newblock \doi{10.4208/cicp.120615.090516a}

\bibitem{Qu2005}
Qu, S., Shastry, V., Curtin, W.A., Miller, R.E.: {A} finite-temperature dynamic
  coupled atomistic/discrete dislocation method.
\newblock Modelling and Simulation in Materials Science and Engineering
  \textbf{13}(7), 1101--1118 (2005).
\newblock \doi{10.1088/0965-0393/13/7/007}

\bibitem{Scholz2001}
Scholz, W., Schrefl, T., Fidler, J.: {M}icromagnetic simulation of thermally
  activated switching in fine particles.
\newblock Journal of Magnetism and Magnetic Materials \textbf{233}(3), 296--304
  (2001).
\newblock \doi{10.1016/S0304-8853(01)00032-4}

\bibitem{Tadmor2011}
Tadmor, E.B., Miller, R.E.: {M}odeling materials.
\newblock Cambridge University Press (2011)

\bibitem{Tranchida2016}
Tranchida, J., Thibaudeau, P., Nicolis, S.: {C}losing the hierarchy for
  non-{M}arkovian magnetization dynamics.
\newblock Physica B \textbf{486}, 57--59 (2016)

\end{thebibliography}
%\bibliography{dek}

\end{document}